\documentclass{aa}  

\usepackage{graphicx}
\usepackage{txfonts}
\usepackage{lipsum}
\usepackage{subcaption}         
\usepackage{lscape}             
\usepackage{placeins}           
\usepackage{longtable}
\usepackage{booktabs}

\usepackage{hyperref}
\hypersetup{
    colorlinks=true,
    linkcolor=blue,
    filecolor=magenta,
    urlcolor=blue,
    citecolor=blue,
    }
    
\begin{document}

   \title{The upper main sequence in h and $\chi$ Persei\thanks{Partially based on observations collected at the Isaac Newton and William Herschel telescopes (La Palma).}}



   \author{A. Marco\inst{1,2}
           , I. Negueruela\inst{3,2},
           N. Castro\inst{4}     
        \and K. Rübke\inst{3}
        }

   \institute{DFISTS, EPS, Universidad de Alicante, Carretera San Vicente del Raspeig s/n,  E-03690, San Vicente del Raspeig, Spain\\
            \email{amparo.marco@ua.es}
            \and Instituto Universitario de Investigación Informática. Universidad de Alicante
            \and Departamento de Física, Facultad de Ciencias, Universidad de Alicante
            \and Leibniz-Institut für Astrophysik Potsdam (AIP), An der Sternwarte 16, D-14482, Potsdam, Germany
            \\}

   \date{Received September 6, 2026 / Accepted October 25, 2026}


\abstract
   {Binary interactions play a crucial role in the evolution and eventual outcomes of massive stars. Rich young open clusters provide large co-eval populations that can be used to constrain the effects of binary evolution. The Double Cluster, $h$ and $\chi$ Persei, with a turn-off mass well above the threshold for core-collapse supernovae, is an excellent test bed for these processes.}
   {We used the observed stellar populations of $h$ and $\chi$ Persei, together with the results of a similar study of NGC 663, to constrain the evolutionary pathways of massive stars and, in particular, the role of binary interaction.} 
   {We present optical spectroscopy of around 80 stars, covering the entire upper main sequence of NGC~869 and most of the brightest B-type stars in NGC~884, and used it to obtain spectral types and derive stellar parameters. We combined this dataset with  Gaia astrometry to evaluate their evolutionary status and possible evolutionary paths.}
  {We find that 40\% of the stars along the photometric sequence of NGC 869 have parallaxes consistent with the cluster mean but proper motions inconsistent at the 2$\sigma$ level, suggesting some sort of dynamical perturbation in this field. We derived atmospheric parameters for almost 60 stars in the upper main sequence (B2\,V and earlier) and found broad agreement between two independent fully automated procedures. We identified the cluster turn-offs between B1\,V and B1.5\,--\,2\,IV and differentiated this feature from the vicinity of the terminal main sequence (TAMS), where we found a population of giants between B1 and B1.5 with a high fraction of Be stars. Three luminous supergiants (HD 14134, HD 14143, HD 13854) are clearly detached from the rest of the population in all our diagrams. }
   {The bulk of the population in all of the massive clusters studied in the Perseus arm is consistent with single-star evolutionary models at a given age, but some B-type supergiants are much more luminous than any other cluster star. Their typical cluster kinematics and stellar parameters suggest a merger origin. Other lower-luminosity B-type supergiants seem consistent with the rest of the population. Kinematic analysis has indicated that most Be stars have not been dynamically perturbed by a supernova explosion. Those below the turn-off are likely to have been formed via case A mass transfer. The more massive Be stars above the turn-off do not display any other characteristic indicating a distinct evolutionary population. }

   
   \maketitle
   \nolinenumbers

\section{Introduction} 

In recent years, it has become increasingly clear that binary interactions play a defining role in the evolution of massive stars and their eventual outcomes \citep[e.g.][]{Marchant2024,Renzo2019}. Large-scale spectroscopic surveys of stars in different environments have revealed chemical signatures consistent with such interactions in a substantial fraction of massive stars \citep[e.g.][]{menon2024,Simon2026,Martinez2026}. Such studies provide very valuable hints about the existence of evolutionary effects, but they are limited by the diverse properties (age, metallicity, environment) of the stars analysed, which in many cases are part of a magnitude-limited sample and chosen because of observability constraints. 

In the Gaia era, rich young open clusters with large and well characterised stellar populations provide a valuable complement to large-scale studies. Their co-eval populations offer a controlled environment in which individual stars that deviate from the bulk of the population can be identified and used to constrain the effects of binary evolution. In this context, a recent paper by \citet{Marco2025} analysed the upper sequence of NGC~663, a rich cluster in the Perseus arm, which was found to have an initial mass on the order of $10^{4}\:\mathrm{M}_{\sun}$. At an age of $\simeq23\:$Ma, the cluster presents a well-defined turn-off, indicated by a substantial population of B2.5\,IV stars, which should have masses around $9\:\mathrm{M}_{\sun}$. There is, however, a rich population of blue supergiants (BSGs), at least four of which appear to have significantly higher masses than the brightest giants near the Hertzsprung gap. This strongly suggests that BSGs are mass gainers and produced after mass transfer in a binary. Interestingly, all of these objects have proper motions typical of cluster members, indicating that they have not been dynamically perturbed by the supernova explosion of their former companions. Given the inferred high masses of the supergiants ($\gtrsim15\:\mathrm{M}_{\sun}$), and the current turn-off mass, this situation seems to favour the scenario in which BSGs are produced via mergers \citep[see][]{menon2024}.

NGC~663 also contains a very large population of Be stars, concentrated around and above the cluster turn-off. In recent years, many indirect lines of argument have suggested that most Be stars form through binary interactions \citep[e.g.][]{Klement2019,bodens2020Be,Dodd2024}, with most population synthesis models suggesting that moderately conservative case B evolution is the predominant channel for forming Be stars \citep[e.g.][]{vinciguerra2020,Rocha2024,Lechien2025}. The population of Be stars in NGC~663 is difficult to reconcile with this scenario, as there is only one Be blue straggler, and no other obvious Be runaways or walkaways that could have been kinematically perturbed by the supernova explosion of the former primary. This could indicate either that stars of 10\,--\,$12\:\mathrm{M}_{\sun}$ starting their lives as primaries in binaries do not frequently undergo supernova explosions or that case A mass transfer from companions below the turn-off is the dominant path to form Be stars. Since the bulk of stars above the turn-off in NGC~663 have masses just above the theoretical limit for supernova explosions \citep[e.g.][]{Jones2014,siess2018}, it is difficult to distinguish between these possibilities.

A younger cluster, with a higher turn-off mass, implying a population of stars well above the threshold for supernova production, could provide a cleaner test of these evolutionary scenarios. The Double Cluster, $h$ and $\chi$ Persei, is particularly well suited for this purpose. Representing the core of the Per~OB1 association, it is one of the richest young open-cluster systems accessible from the Northern Hemisphere and is subject to a particularly low extinction. Consequently, it has been extensively studied, with a substantial body of observational work predating the Gaia era, including photometric and spectroscopic investigations \citep[e.g.][]{Johnson1955,wildey1964,schild1965,waelkens1990}. \citet{Currie2010} carried out the first extensive study of their stellar population, down to subsolar masses using optical spectroscopy and optical/infrared photometry. They determined an age of $\approx14$ Ma and concluded that, except for mass and spatial distribution, the clusters are nearly identical in every measurable way.

A number of recent studies have used Gaia astrometric data (positions, proper motions,  and parallaxes) to determine cluster parameters of open clusters by using automated techniques. Several catalogues include $h$ \&  $\chi$~Persei, analysed either with Gaia DR2 data \citep{Cantat2020, dias2021} or, more recently, DR3 data \citep{Hunt2023}.The latter provides the most recent estimate of global parameters for both clusters (see Table~\ref{tab:parametersHunt} and section~\ref{gaia}).

In this paper, we study $h$ (NGC~869) and $\chi$~Per (NGC~884), by presenting optical spectroscopy of around $80$ stars that include the whole upper main sequence of NGC~869 and most of the brightest B-type stars in NGC~884, and obtaining spectral types and stellar parameters. The combination of this dataset with the corresponding Gaia astrometry allows for a thorough investigation of the stellar population and the placement of strong constraints on the evolutionary pathways involved in their creation.

The paper is organised as follows. In Section~\ref{sec:data}, we describe the observations and reduction procedures used; in Section~\ref{sec:results}, we characterise the stellar population by determining spectral types and stellar atmospheric parameters; in Section~\ref{sec:discussion}, we analyse the upper main sequence population and compare observed properties to current theoretical binary evolution channels. Finally, Section~\ref{sec:conclus} provides a summary of our main conclusions. Throughout the paper, we use the numbering system used in Simbad, which follows \citet{Oosterhoff1937}.

\section{Observations and data}
\label{sec:data}

We obtained low-medium resolution spectroscopy for around 80 stars in the two clusters with two different instruments, the Intermediate Dispersion Spectrograph (IDS) on the 2.54-m Isaac Newton Telescope (INT), and the Intermediate dispersion Spectrograph and Imaging System (ISIS) on the 4.2-m William Herschel Telescope (WHT), both placed  at the La Palma observatory. Additionally, we made use of Gaia DR3 data corresponding to the fields of NGC~869 and NGC~884 \citep{GaiaCollaboration}.

\subsection{Spectroscopy}
\label{sec:spectra}

The sample observed was selected before accurate proper motions from Gaia were available. It consists of all the stars whose colours and magnitudes suggested that they were upper main sequence members of NGC~869. This sample is intended to be complete down to B2\,V, (and a few fainter stars in the cluster core that serendipitously fell in the slits), plus most of the brightest B-type stars in NGC~884.

Fifteen stars in NGC~869 were observed with the IDS spectrograph, equipped with the 235-mm camera and the EEV13 CCD. We used the R900V grating (covering  3700\,--\,5500\,\AA), and we set a $1.0 \arcsec$ slit width, obtaining a resolving power of $R\sim3\,600$. The observations were taken on the nights of 2003 July 3 and 8.

Four stars in NGC~869 were observed with the ISIS spectrograph in both the blue and red arms, equipped with the EEV12 CCD and MARCONI2 CCDs, respectively. We used the R1200B and R1200R gratings (covering  3900\,--\,4700\AA\, and 6000\,--\,6700\AA\,), and we set a $1.0 \arcsec$ slit width, obtaining  a resolving power of $R\sim5\,300$. The observations were obtained on the night of 2003 August 17. Two further supergiants associated with the cluster, which have traditionally been used as MK standards, were observed with the same setup, but only in the blue arm, on the night of 2002 November 21.

Forty-seven stars in and around NGC~869 were observed with the ISIS spectrograph, on the night of 2003 November 15. Only the blue arm was used, equipped with the R600B grating (covering  3600\,--\,5300\,\AA) and the EEV12 CCD. We set a $1.0 \arcsec$ slit width, obtaining a resolving power of $R\sim2\,300$. Seventeen stars in NGC~884 were observed with the ISIS spectrograph on the night of 2005 September 5, using exactly the same setup. 

None of these runs included a sufficient number of arcs to measure radial velocities, because the main aim of the observations was to derive spectral classification.  All the long-slit spectroscopic data were reduced  with  the \textit{Starlink} packages {\sc CCDPACK} \citep{Draper2000} and {\sc FIGARO} \citep{Shortridge1997}, and analysed using {\sc FIGARO} and {\sc DIPSO} \citep{Howarth1998}.

\begin{figure*}
\sidecaption
\includegraphics[width=12cm]{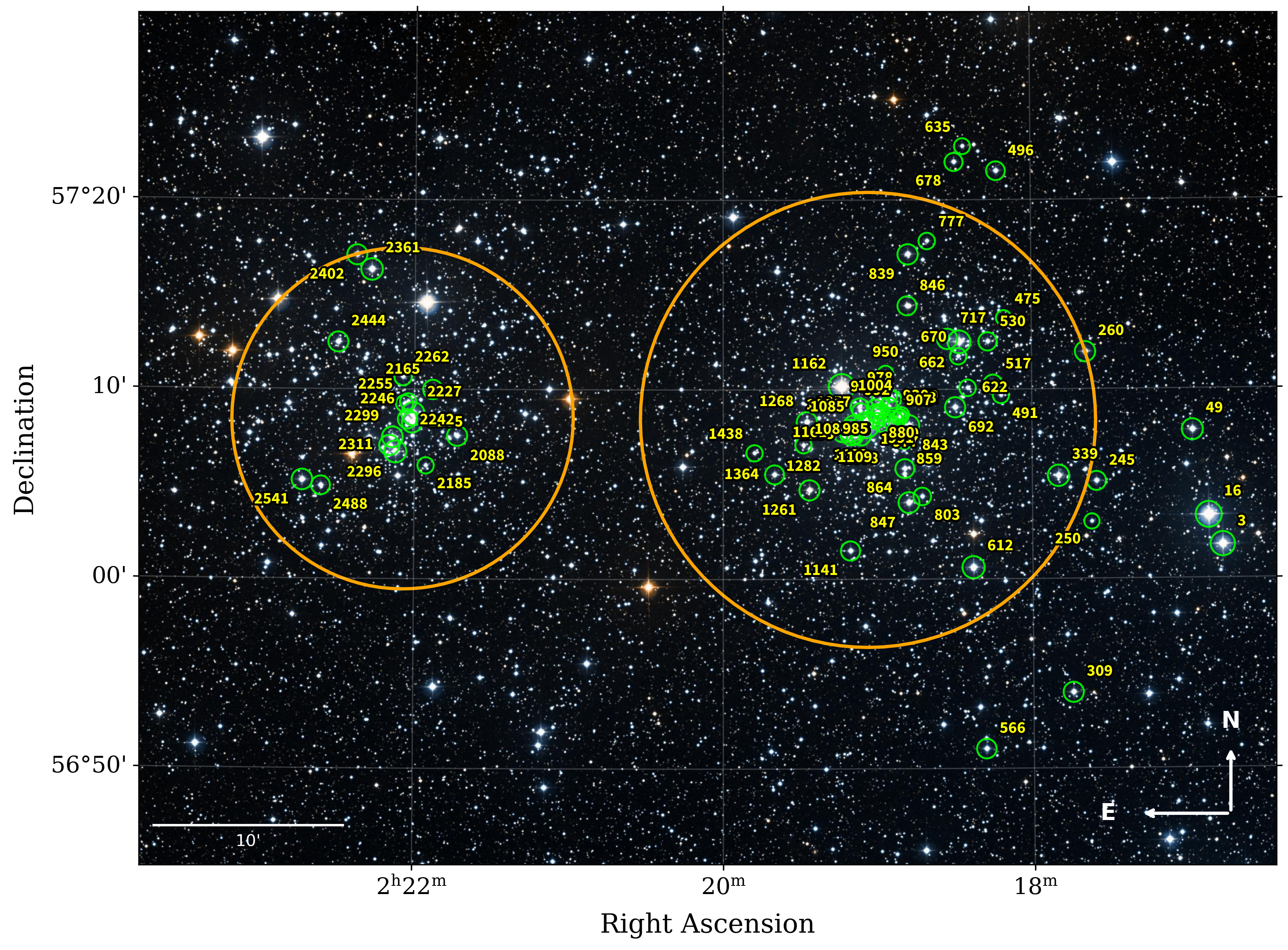}
\caption{Spatial distribution of our sample, with the Oosterhoff numbers of all the stars observed. The large circles indicate the tidal radii calculated for both clusters by \citet{Hunt2023}. Insets of the two cluster cores can be found in Fig.~\ref{fig:coremaps}.
\label{fig:bigmap}}
\end{figure*}   

The list of stars observed in the area around NGC~869 can be found in Table~\ref{tab:param869}, while those around NGC~884 are listed in Table~\ref{tab:param884}, together with their spectral types and parameters derived from the analysis (see next Section). The parameters for four BSGs associated with NGC~869 have been taken from literature. They are all MK standards \citep{Negueruela2024} and have had their parameters derived from high-resolution spectra \citep{deBurgos2024}. We list them in Table~\ref{tab:BSGs869}. The spatial distribution of the stars observed can be seen in Fig.~\ref{fig:bigmap}.

\begin{table*}[ht]
\caption{Spectral types and stellar parameters for stars in NGC~884.}
\label{tab:param884}
\centering
\begin{tabular}{l l l c c c }
\hline\hline
\noalign{\smallskip}
Oo &
Other &
Spectral &
 $v\,\sin\,i$ &
$T_{\mathrm{eff}}$ &
 $\log\,g$\\
 Number&
 Name&
Type &
(km\,s$^{-1}$) &
(K) &
(dex) \\
\noalign{\smallskip}
\hline
\hline
\noalign{\smallskip}
2361&HD 14476	     &       B1\,II    &       20      &       $26000^{+1000}_{-1900}$    &       $3.45^{+0.14}_{-0.15}$    \\
2299&BD +56 575\tablefootmark{a}	     &       B1\,II-III	     &       100     &       $25000^{+1400}_{-1300}$    &       $3.49^{+0.19}_{-0.12}$    \\
2402&BD +56 582      &       B1\,IIIe  &         &  Strong emission       &           \\
2235&BD +56 571       &       B1\,IIIn  &       160     &       $24000^{+1000}_{-1600}$    &       $3.42^{+0.10}_{-0.14}$    \\
2444&BD +56 584      &       B1\,V     &       120     &       $24000^{+2000}_{-1200}$    &       $3.80^{+0.19}_{-0.15}$    \\
2488&BD +56 586      &       B1\,V     &       140     &       $24000^{+1300}_{-1700}$    &       $3.52^{+0.14}_{-0.15}$    \\
2165&BD +56 566      &       B1\,Ve    &         &       Strong emission    &    
\\
2246&BD +56 572\tablefootmark{b} &       B1\,V\tablefootmark{c}  &       120     &       $24000^{+3000}_{-1600}$    &       $3.7^{+0.3}_{-0.2}$     \\
2227&HD 14443	     &       BC1.5\,Ib       &       20      &       $22000^{+1600}_{-1900}$    &       $3.0^{+0.1}_{-0.2}$     \\
2296&BD +56 574      &       B1.5\,Ibn	     &       160     &       $22000^{+1400}_{-1000}$    &       $3.06^{+0.10}_{-0.11}$    \\
2088&BD +56 563      &       B1.5\,IIIe	     &           &     Strong emission      &          \\
2311&BD +56 576\tablefootmark{d}	     &       B1.5\,III\tablefootmark{e}	     &       20      &       $24000^{+1000}_{-1800}$    &  $3.52^{+0.14}_{-0.17}$    \\
2541&HD 14520	     &       B1.5\,III &       60      &       $23000^{+1600}_{-1300}$    &       $3.2^{+0.2}_{-0.1}$     \\
2262&EM* MWC 39      &       B1.5\,Vn\tablefootmark{f}   &       260     &       $25000^{+2000}_{-3000}$    &       $4.00^{+0.13}_{-0.30}$     \\
2255 &       &B2\,V     &       360     &       $21000^{+2000}_{-1500}$    &       $3.5^{+0.2}_{-0.1}$     \\
2185&	     &       B2.5\,V   &          &  Poor spectrum         &          \\
2242&	     &       B2.5:\,V:shell\tablefootmark{g}    &         & Poor spectrum       &      \\
\noalign{\smallskip}
\hline
\hline
\end{tabular}
\tablefoot{Stellar parameters based on the global fit method. Characteristics of individual stars:\\
\tablefoottext{a}{V595 Per};
\tablefoottext{b}{V619 Per};
\tablefoottext{c}{$\beta$ Cepheid};
\tablefoottext{d}{V621 Per};
\tablefoottext{e}{Eclipsing binary};
\tablefoottext{f}{Known Be, but not obvious in blue spectrum};
\tablefoottext{g}{Likely binary}
}
\end{table*}
\begin{table*}
\caption{Spectral types and stellar parameters for four luminous BSGs around NGC~869.}
\label{tab:BSGs869}
\centering
\begin{tabular}{l l l c c c }
\hline\hline
\noalign{\smallskip}
Oo &
Other &
Spectral &
 $v\,\sin\,i$ &
$T_{\mathrm{eff}}$ &
 $\log\,g$\\
Number &
Name &
Type &
(km\,s$^{-1}$) &
(K) &
(dex) \\
\noalign{\smallskip}
\hline
\hline
\noalign{\smallskip}
 1162&HD 14143	     &       B2\,Ia    &       55      &       $19900^{+500}_{-300}$    &       $2.50^{+0.04}_{-0.07}$    \\
 1057&HD 14134     &       B3\,Ia	     &      37     &       $16700^{+400}_{-300}$    &       $2.17^{+0.05}_{-0.05}$    \\
   16&HD 13854       &       B1\,Iab  &       55     &       $22500^{+300}_{-300}$    &       $2.64^{+0.04}_{-0.04}$    \\
    3&HD 13841      &       B1.5\,Ib    &        43    &       $22400^{+600}_{-500}$    &       $2.76^{+0.06}_{-0.06}$    \\
\noalign{\smallskip}
\hline
\hline 
\end{tabular}
\tablefoot{Spectral types and stellar parameters taken from \citet{deBurgos2024} }
\end{table*}

Unlike the sample analysed by \citet{Marsh2012}, which used a multi-object spectrograph, we concentrated on the cluster cores. The high coverage of the core regions can be best appreciated in Fig.~\ref{fig:coremaps}. Our sample is almost complete 
for bright stars ($V\lesssim 11$, roughly corresponding to B2\,V), especially across the face of NGC~869, allowing us to explore the characteristics of all stars around and above the turn-off. Contrarily, we only have a few later-type stars, which were much better represented in the work of \citet{Marsh2012}, but sampled primarily in the surrounding association. The only bright star that we are missing is 141 (BD +56 528), which lies to the south of the cluster core, and has recently been classified as B1\,Vn \citep[][]{deBurgos2023} 

Our coverage of NGC~884 is not so complete, although the number of bright stars is much smaller. Nevertheless, there are only two bright stars in the core of NGC 884 that we did not observe (apart from the red supergiant RS~Per). The first is 2284 (BD +56 573), an extreme Be whose spectrum can be found in \citet{Steele1999}. Its spectral type is difficult to assess due to heavy veiling, but it is not far from B1\,IV. The second is 2371 (BD +56 578), which was classified BN2\,II-III in \citet{deBurgos2020}. 

In summary, we cover the whole area of the double cluster down to B2\,V. We are only missing a handful of stars in the area between the two clusters and a few likely members of NGC~884 at large distances from its core.

\section{Results}
\label{sec:results}

\subsection{Spectral classification}

\begin{figure*}[h!]
   \centering
    \includegraphics[width=\columnwidth]{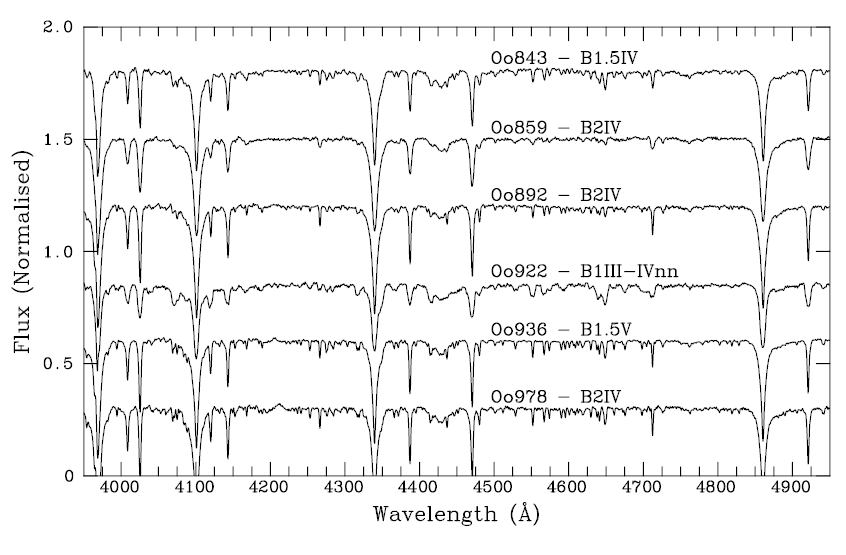}
    \includegraphics[width=\columnwidth]{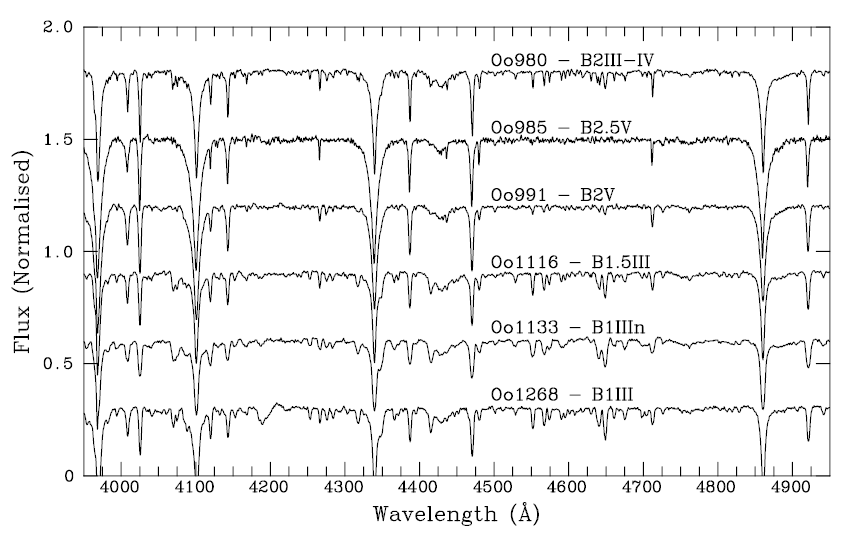}
      \caption{Blue spectra of a representative sample of stars in the core region of NGC~869. The very fast rotator 922 (BD +56 515) is given as a Be star in SIMBAD, but is not listed as such in any reference that we have checked. Given its very shallow lines, it is likely an unresolved binary.} 
         \label{fig:core}
   \end{figure*}

Spectral classification was carried out by direct comparison of our spectra to the new grid of MK standard stars presented in \citet{Negueruela2024}. Although most of the spectra have somewhat lower resolution than the standards, the difference is sufficiently small to allow direct application of the criteria. Moreover, several of the brighter stars observed are classification standards in the new system. Our sample is so large that internal consistency can be achieved by repeated comparison between spectra of similar spectral types. This allows the classification of many objects as intermediate between luminosity classes.  Moreover, consistency was also achieved with the sample of \citet{deBurgos2020}, who observed the brightest stars in the wider Per OB1 association, including some of the most luminous objects in our sample\footnote{The only difference between the two samples is the classification of 339 (HD 13969), which we move from B0.7\,III to B1\,III, and 2296 (BD +56 574), which shifts from B1.5Ib-II to B1.5Ibn. These are minimal changes, attributable to the difference in resolution.}. As an example, Fig.~\ref{fig:core} shows a representative sample of stars in the core region of NGC~869. Other spectra are displayed in Figures~\ref{fig:luminous} to~\ref{fig:to_chi}.

Our classifications assume that all stars are single objects. This is, of course, incorrect, as some of the stars are known to be binaries. A notable example is 2311 (BD +56 576 = V621 Per), which is a well-know eclipsing binary. The secondary, however, is not detected in the spectra, and our classification refers to the brighter member, which \citet{Southworth2004} assumed to be the primary.  None of the spectra that we have analysed is clearly a double-lined spectroscopic binary at this resolution, although some offer signs of binarity. For example, 496 (BD +56 488) presents very asymmetric absorption lines, and almost certainly has a later-type companion. Another object that is a strong SB2 candidate is 530. Both stars were excluded from the analysis. Other objects have subtler, but also strong signs of binary companions, such as 922 (BD +56 515 = V613~Per) in Fig.~\ref{fig:core}, whose lines are much shallower than those of stars of similar spectral type in the same plot.

With the strict application of the classification criteria, there is a very good correlation between luminosity class and brightness, which was not the case for \citet{Slesnick2002}. There are, however, a few remarkable exceptions. The star 843 (BD +56 510) is far too bright ($G=9.3$) for its spectral type, B1.5\,IV. \citet{Marsh2012} derive a  $\log\,g = 3.4$, more typical of a giant. This suggests a misidentification for our spectrum or an unresolved SB2. Conversely, 803 (BD +56 505) at G = 10.4 is far too faint for its spectral type (B1\,III-IV). However, \citet{Marsh2012} derive stellar parameters fully compatible with ours, with a $\log\,g \approx3.5$, typical of a giant, making a misidentification unlikely. Similarly, 678 (LS V +57 9) is quite faint ($G=10.2$) for its spectral type (B1.5\,IV), but again \citet{Marsh2012} gives stellar parameters fully compatible with ours and appropriate for the type.

\subsection{Astrometric membership analysis}
\label{gaia}

Using Gaia data, several authors have identified cluster members utilising different Bayesian clustering algorithms. These methods find statistical significant groupings in both spatial coordinates and astrometric parameter space (proper motions and parallax). Most recently, \citet{Hunt2023} used the Gaia DR3 catalogue and the HDBSCAN algorithm to assign membership to a large sample of clusters and to derive their mean astrometric parameters. Mean proper motion and parallax were calculated from stars classified as members within the estimated tidal radius of each cluster. Additionally, maximum likelihood cluster distances were derived using the method of \citet{Cantat-Gaudin2018}, after applying the parallax zero-point correction of  \citet{Lindegren2021}. Their results for NGC~869 and NGC~884 are given in Table~\ref{tab:parametersHunt}.

\begin{figure*}
\sidecaption
\includegraphics[width=6cm]{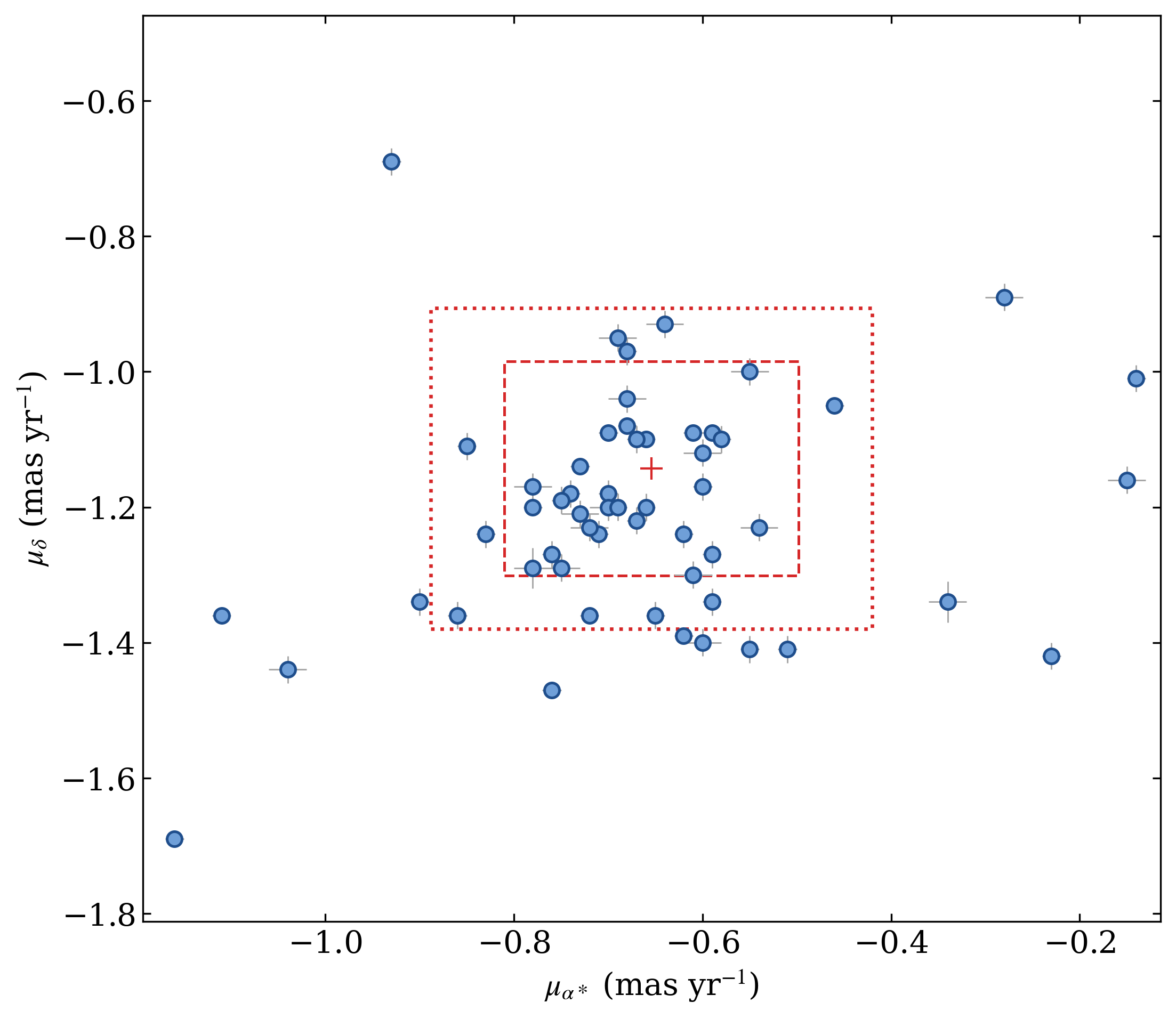}
\includegraphics[width=6cm]{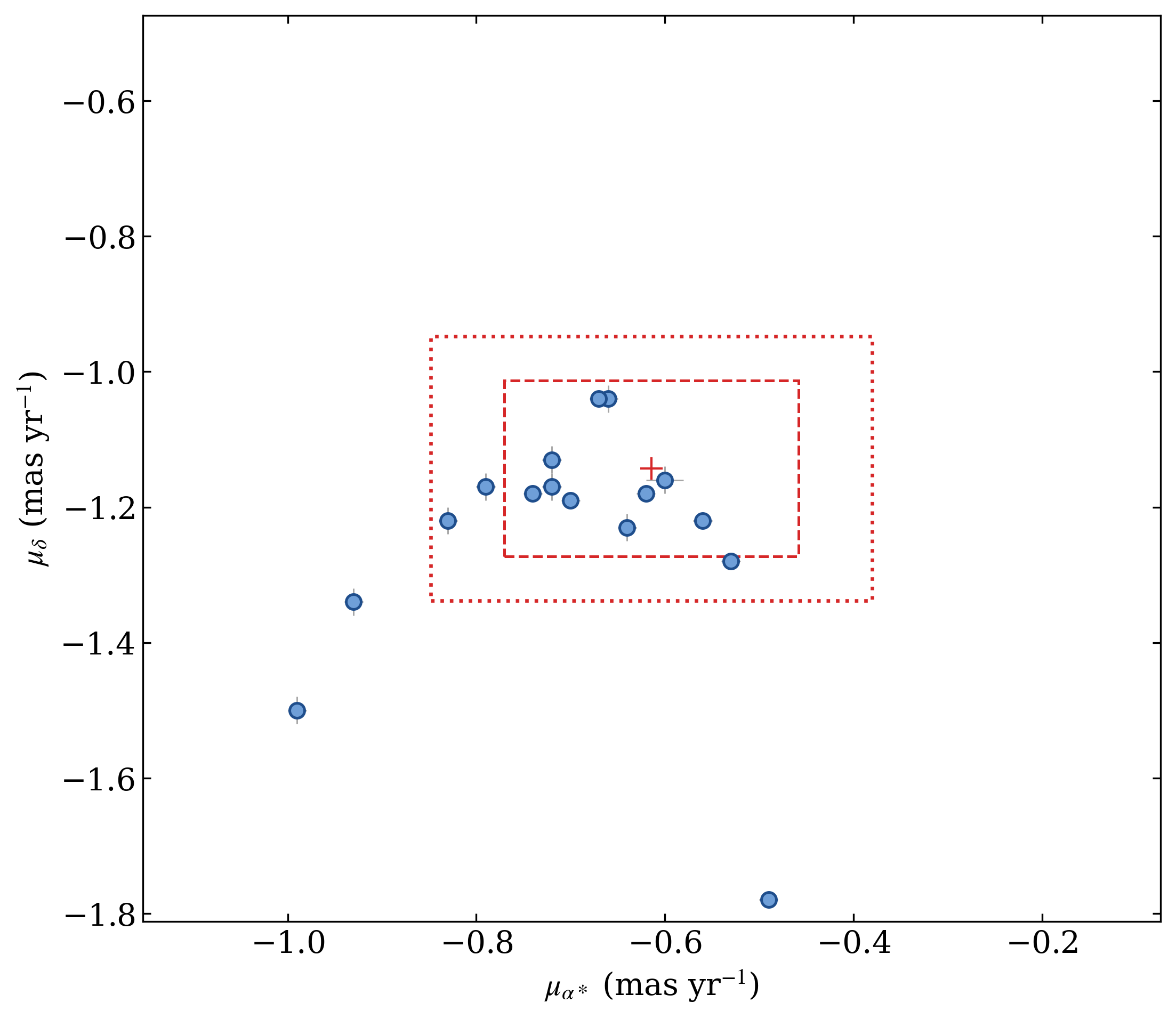}
\caption{Proper motions for the stars in our sample. Only photometric members of NGC~869 (left) or NGC~884 (right) with RUWE$<1.4$ and a parallax compatible with cluster membership have been included. The red crosses mark the average values given by \citet[see Table~\ref{tab:parametersHunt}]{Hunt2023}; with the two rectangles indicating two and three times the standard deviation ($\sigma$) of cluster members selected by \citet{Hunt2023}.} 
\label{fig:propermotions}
\end{figure*}   

The astrometric parameters derived for the two clusters are essentially indistinguishable: their mean $\varpi$ and pmDEC are identical, while the mean pmRA are consistent within the corresponding dispersions. It is thus difficult to assign individual stars in the region of overlap between the two clusters to either cluster. The adopted distances, defined as the 16th percentiles of the maximum-likelihood distance distributions, are $2273\pm15$~pc for NGC~869 and $2286\pm15$~pc for NGC~884, placing the two clusters at essentially the same distance within the uncertainties. To assess the astrometric membership of individual stars, we adopt the standard deviations ($\sigma$) of the distributions of the astrometric parameters of the members listed \citet{Hunt2023}.

Basic Gaia DR3 data ($G$ magnitude, parallax and proper motions) for our target sample in NGC~869 and NGC~884 are displayed in Table~\ref{tab:gaia869} and Table~\ref{tab:gaia884}, respectively. Inspection of these tables reveals an unexpectedly large number of targets selected from photometric membership lists whose astrometric parameters are inconsistent with the mean cluster values listed in Table~\ref{tab:parametersHunt} at 1 $\sigma$. This is also the case when we restrict ourselves to stars located within the high-density regions around the cluster centres. This sharp contrast between the photometric membership and the astrometric membership was not observed for NGC~663 \citep{Marco2025}, nor for older clusters \citep{Quintana2025}. A quick inspection of other massive clusters in the Perseus arm, such as NGC~457, NGC~4719 or NGC~7510, indicates a substantially smaller fraction of photometrically selected members with astrometric parameters inconsistent with cluster membership.

Only two stars in our sample have parallaxes that definitely make them non-members, Oo~264 (B8\,V; $\varpi=0.89\pm0.02$~mas) and 649 (B3\,IV; $\varpi=0.69\pm0.03$~mas). The rest are consistent with the average parallax. Some objects are almost 3$\sigma$ away, but the errors in the parallaxes of bright stars are known to be underestimated in DR3 \citep{maiz2021}. A few have very poor astrometric parameters, and their membership cannot be ascertained from DR3 astrometry. A significant fraction have parallaxes consistent with cluster membership but proper motions that differ significantly from the cluster means. This is illustrated in Fig.~\ref{fig:propermotions}, where we compare the distribution of proper motions for the sample consistent in parallax with the astrometric mean values given by \citet{Hunt2023}. Since our sample only contains bright stars, we quantified this effect for fainter targets: we adopted the cluster centre given by \citet{Hunt2023} for NGC~869 (34.7655108307$^\circ$, 57.1401175113$^\circ$), and retrieved the DR3 data for all stars within the cluster tidal radius ($0.2^\circ$). We then applied 2-$\sigma$ cuts to proper motions and parallax and plotted the standard CMD, $G$.vs.($BP-RP$), for the stars satisfying these criteria. After this, we selected objects along the photometric sequence defined by the members according to \citet{Hunt2023}. 

For stars along this sequence with $G<14$, 224 objects have parallaxes consistent with the cluster at the 2$\sigma$ level, but only 130 of them (58\%) have proper motions consistent with the cluster. Extending the selection to $G<16$ yields 630 stars with consistent parallaxes, of which 391 (62\%) also satisfy the proper-motion criterion. For NGC 884, when carrying out the same analysis, but now using its tidal radius of 9\arcmin, the numbers are 91/109 (83\%) for $G<14$, and 238/290 (82\%) for $G<16$. The discrepancy is therefore substantially more pronounced in NGC 869. Other clusters examined show considerably smaller differences between photometric + parallax and proper motion membership.  When we examine the spatial location of stars failing the proper motion criterion, we find that they are evenly distributed throughout the field.

\begin{figure*}
\sidecaption
\includegraphics[width=6cm]{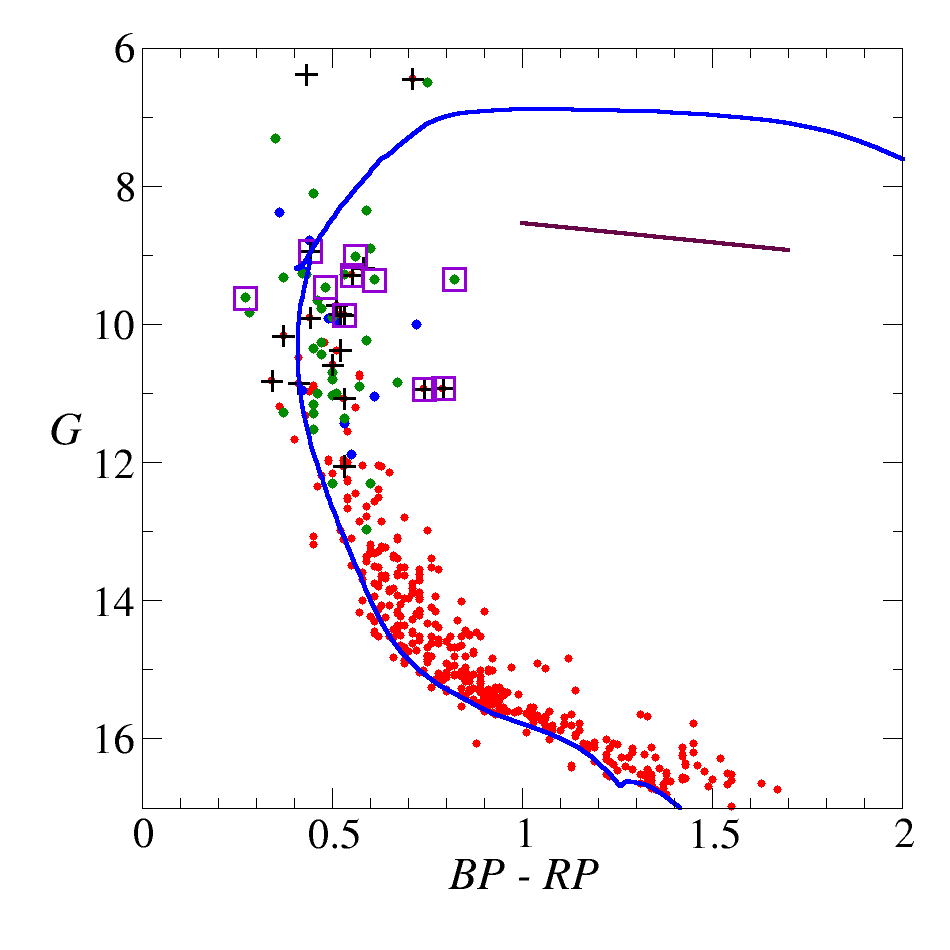}
\includegraphics[width=6cm]{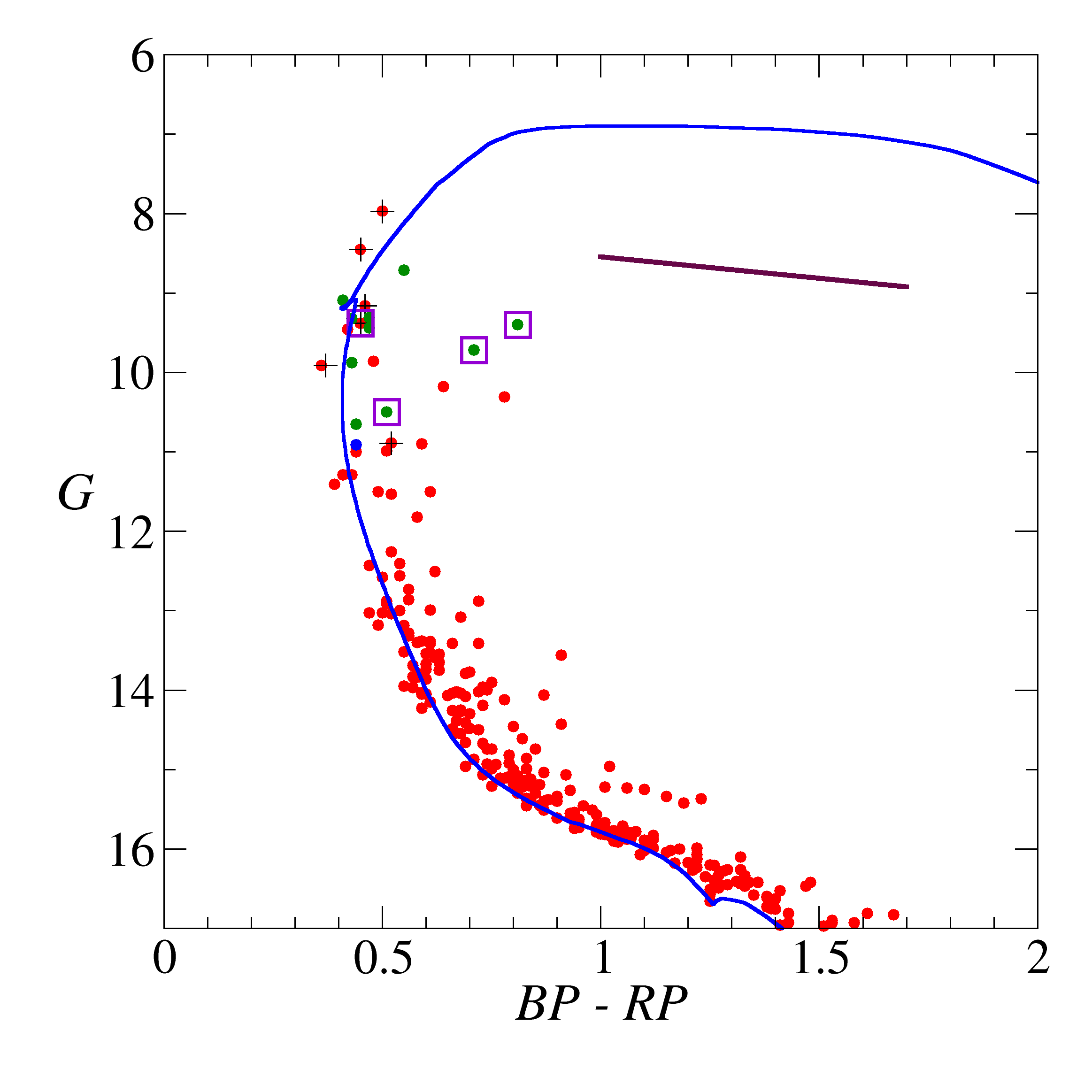}
\caption{Left: CMD ($G$ vs. ($BP-RP$)) for stars in NGC~869. Red dots are members from \citet{Hunt2023}. The remaining symbols correspond to stars observed spectroscopically in this study: blue circles are stars with RUWE > 1.4, green circles are non-astrometric members, black crosses are astrometric members, and violet open squares are Be stars. The blue line displayed is a PARSEC 1.2 \citep{Bressan2012} $\log t= 7.26$ isochrone at solar metallicity reproducing distance and average reddening as estimated by \citet{Hunt2023}, with no attempt at fitting the data. The maroon segment represents the reddening vector. 
Right: Same but for stars in NGC~884, with the same isochrone. }
\label{fig:Gaiadiagrams}
\end{figure*}     

We have to conclude that there is a large population of objects that are photometric members and are at the same distance as the clusters, but have different kinematics. In light of these results, there must exist a strong dynamical disturbance affecting the clusters, especially NGC~869. Possible explanations for this anomaly include the following:
\begin{itemize}
\item The core of NGC~869 is heavily contaminated by members of the surrounding association with different kinematics. \citet{deBurgos2020} conclude that the population of the wider association has a broader distribution of proper motions than those of the clusters. But this possibility does not seem very likely, as it would imply a strong over-density of association members projected on top of the cluster core, and decreasing with distance to the core. Conversely, \citet{deBurgos2020} place the expansion centre of the association one degree north of the clusters. 

\item There is a strong gravitational interaction between the two clusters, leading to perturbed internal kinematics. This possibility can almost certainly be rejected, because, given the lower mass of NGC~884, the effect should be stronger in its core.

\item Stellar trajectories at the centre of NGC~869 have been perturbed, with many stars affected by recent interactions or in the process of being ejected because of a supernova explosion in a binary. This is not seen in clusters with similar masses and ages, and should result in a moderately large population moving away from the cluster, which is not observed. Again, \citet{deBurgos2020} conclude that the spatial distribution of proper motions in the
extended population of blue and red supergiants in Per~OB1 does
not follow an expanding structure centred on the h and $\chi$ Persei
double cluster. In addition, the proper motion vectors of four runaways O-type stars do not point outwards from h and $\chi$ Persei, but to a far more extended region. 

\item The two high-density clumps within NGC~869 represent two clusters in the process of merging, which leads to perturbed kinematics. This possibility cannot be excluded, but \citet{deBurgos2020} conclude that no important differences are detected in
the distribution of parallaxes, proper motions, and RVs when
stars in h and $\chi$ Persei or the rest of the association are considered, which
suggests a very extended dynamically interrelated population.
\end{itemize}

Whatever the cause, if we relax slightly the constraints, the situation becomes much more satisfactory. Almost all the stars in NGC~884, as well as the vast majority of red supergiants and BSGs in the halo surrounding the cluster have proper motions within 3$\sigma$ of the cluster means. For NGC~869, the situation is a bit worse, but many of the stars outside the 3$\sigma$ limit are located at large distances from the core, and may be identified as part of the broader association. In Fig.~\ref{fig:Gaiadiagrams}, we show the Gaia CMDs for the two clusters. Red circles represent the 403 and 287 members listed in \citet{Hunt2023} for NGC~869 and NGC~884 (Table~\ref{tab:parametersHunt}). The rest of the symbols identify the stars in our sample: blue circles are stars with RUWE $> 1.4$ which do not satisfy the astrometric quality criterion of \citet{Lindegren2021} and are left out by \citet{Hunt2023}; green circles are stars that \citet{Hunt2023} do not label as members; black crosses are astrometric members according to \citet{Hunt2023}, and violet open squares identify Be stars.

We observe a narrow sequence for faint stars that widens considerably between $G=14$ and 15, the locus of late-B and early-A stars, where the effects of diverse initial rotational velocities become obvious. Around $G=11.5$, coinciding with the limit of our sample, the sequence becomes even broader. This is partly due to many Be stars having redder colours (those observed by us are indicated), but also by an extension to the blue. This extra widening cannot be due to differential reddening, as the area covered is almost the same as that containing the photometric members, but must reflect a distribution of stellar properties. Despite the much closer agreement between photometric and proper motion membership, NGC~884 contains many fewer members, as it is a smaller cluster. 

\subsection{The Be star content}
\label{sec:bes}

There is a fair number of Be stars catalogued in the vicinity of the Double Cluster. We have observed a few at some distance from the clusters, such as 49 (BD +56 473) or 566 (BD +56 493), and all those catalogued in the central region of NGC~869\footnote{Note that \citet{Marsh2012} list the B3\,Ia supergiant HD~14134 (1057) as a Be star.}. As in other clusters, Be stars concentrate around and above the turn-off. The fraction of Be stars is moderately high. Of the $\approx45$ early-B stars that we observed within the tidal radius of NGC~869 \citep[as defined by][]{Hunt2024}, nine are Be stars. Representative spectra are shown in Fig.~\ref{fig:bes}.

The less luminous Be stars observed have spectral types B2\,V. The only exception is 2244, in NGC~884, which we classify B2.5:\,Ve, but this is a poor spectrum with shell characteristics and evidence of a second component, which render this classification uncertain. There are three fainter stars classified as Be stars from photometry in the field of NGC~869\footnote{\citet{Keller2001} claim a fourth Be star, their star 191. However, this is almost certainly a mistake, as both their stars 156 and 191 coincide with Oo~1438, which is only one star in DR3.}, and three in the field of NGC~884, two of them confirmed in prism surveys \citep{Keller2001}. The list of all the confirmed Be stars within the tidal radii of the two clusters is presented in Table~\ref{tab:bestars}.

Many of the Be stars show weak emission, and it has been possible to analyse them as normal stars. Others have stronger features, which distort the global fit, or do not allow a good determination of line wings. Because of this, we have not included in the analysis stars 49, 309, 1161, 1261, 1282, and 1438 for NGC~869, and 2088, 2165, 2244 and 2402 for NGC~884. Some of these stars are displayed in Fig.~\ref{fig:bes}. An interesting case is 517 (BD +56 489). This object shows weak absorption metallic lines suggestive of a shell spectrum. Based on the broad absorption features, if assumed photospheric, we classify it as B3\,III, which is later than any other giant. We cannot discard the possibility of an underlying hotter spectrum that is masked by the shell component. This can be elucidated with a higher-resolution spectrum. We have not derived parameters for this star, either.

For the rest of the Be stars, we have derived parameters (as discussed in the next section) assuming that their lines do not display contamination from the circumstellar disk. In general, their parameters are consistent with those of the other stars of the same spectral type. However, fast rotation may sometimes lead to an underestimation of the effective gravity. This is discussed in the next section.

\subsection{Stellar atmospheric parameters}
\label{sec:params}

\begin{figure*}
\resizebox{\textwidth}{!}{\includegraphics[angle=0]{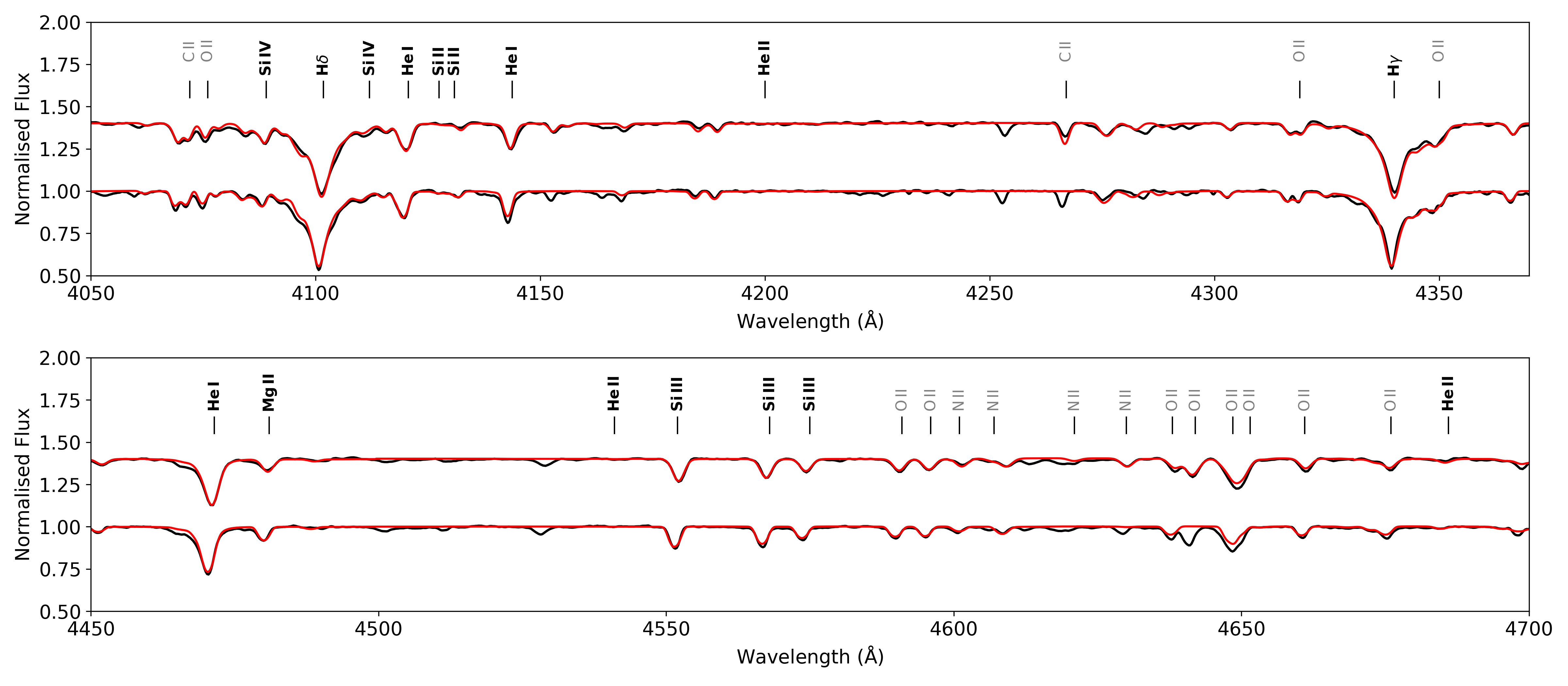}}
\centering
\caption{Model spectra (red) fitted to the spectra of two stars: The top line is HD~13969 with R$\sim$2300, and the bottom line is BD~$+56\degr$ 511 with R$\sim$5300. Top panel: Region including the H$\delta$ and H$\gamma$ lines, whose wings determine the effective gravity. Bottom panel: Region containing the majority of metallic lines. The diagnostic lines used in the fit are marked on top of the spectrum of HD~13969 as black labels. C, N, O lines that are generally utilised for abundance determination are indicated with grey labels, but we did not use them in this study. \label{fig:bertofits}} 
\end{figure*}

To derive stellar atmospheric parameters, we followed the method described by \citet{castro2012}, also used by \citet{Marco2025}. This procedure compares the observed spectrum with a grid of synthetic stellar atmosphere models generated with the {\sc fastwind} code \citep{santolaya1997,puls2005}, as detailed in \citet{castro2012}. An automated, $\chi^2$-based algorithm identifies the model spectrum that best reproduces key hydrogen, helium, and silicon lines in the 4000\,--\,5000\,$\AA$ range. The list of lines employed in the analysis can be found in table~1 of \citet{castro2012}. This grid is appropriate for giant and supergiant B stars, which show moderately strong metallic lines. For higher-gravity stars, which generally display only a few weak metallic lines (particularly for types later than B1.5), we instead used the auxiliary grid of models, described in \citet{Marco2025}, which extends the coverage to higher gravities and temperatures and assumes solar abundances for all the elements.

Since most of the stars have been observed at relatively low spectral resolution, we do not calculate $v \sin\,i$, but rather follow an iterative process to estimate rotational broadening. Initially, we assumed a default value of $20\:$km\,s$^{-1}$ and compared the resulting best-fit model with the observed spectrum. For slow rotators, where the observed line broadening is dominated by the instrumental profile, this should provide a good fit. If this is insufficient to reproduce the observed line profiles, we increased $v \sin\,i$ and repeated the analysis. We iterated until no significant improvement was found by further increasing the rotational velocity, obtaining a rough estimate of the total broadening. When a high value is inferred, it should simply be interpreted to indicate the presence of significant rotational broadening, and not a quantitative measurement of $v \sin\,i$. 

Example fits to two spectra, corresponding to the highest and lowest resolution used, are displayed in Fig.~\ref{fig:bertofits}. The parameters derived are listed in Table~\ref{tab:param884} for stars in NGC~884, and Table~\ref{tab:param869} for stars in NGC~869. The quoted uncertainties represent the range of models that can be considered compatible with the observed spectrum (at a minimum, the grid step). In general, the parameters derived for non-emission stars are to a large degree consistent with the spectral types and luminosity classes assigned. The fits to the mild Be stars are, in most cases, poorer. In general, rapid rotators. whether Be or not, tend to be given lower gravities than corresponds to their luminosity class, and this drives the fit towards lower temperatures, as well. Despite this limitation, the bulk of the values obtained are internally consistent, even if the parameters for individual stars should be treated with caution.

We did not derive parameters for stars with spectral types B3\,V and later. These generally have the lowest-quality spectra, as the stars are fainter and were often observed serendipitously when falling within the slit used for a brighter target. Moreover, extensive testing showed that, in the absence of strong metallic lines, the complex dependence of H and \ion{He}{i} lines on both $T_{{\rm eff}}$ and $\log\,g$ in mid-B types, can cause the global fit not to converge to the correct parameters.

Large spectroscopic surveys such as WEAVE \citep{Jin2024} and 4MOST \citep{deJong2019} will soon provide very large samples of moderate-resolution spectra of OB stars, making automated and robust parameter determination increasingly important. As an independent check on our results and to assess the limitations of the procedure, we used the recently developed \textit{HiLineThere} algorithm within the Astro+ database \citep{Rubke2026} to analyse a subset of the spectra. Unlike the global-fitting procedure adopted here, Astro+ utilises individual diagnostic lines and calculates values for $v \sin\,i$ and macroturbulent broadening. It therefore requires a minimum spectral resolution around $R\approx3\,000$. We consequently restricted the comparison to the subset of NGC~869 spectra obtained at $R=3\,600$ or $R=5\,300$. 

\begin{table*}
\caption{Spectral types and stellar parameters for stars in NGC~869.}
\label{tab:param869ASTRO}
\centering
\begin{tabular}{lllccc}
\hline
\hline
\noalign{\smallskip}
Oo &
Other &
Spectral &
 $v\,\sin\,i$ &
$T_{\mathrm{eff}}$ &
 $\log\,g$\\
Number  &
Name &
Type &
(km\,s$^{-1}$) &
(K) &
 (dex)\\
 \noalign{\smallskip}
\hline
\hline
\noalign{\smallskip}
612 &BD+56\,\,498\tablefootmark{a}     &   B0.7\,II&$143\pm90$	 &$24000\pm1000$&	 $3.58\pm0.08$  \\
692 &BD+56\,\,501      &   B0.7\,III	 &$196\pm13$&  $25400\pm500$   &	 $3.5\pm0.1$  \\
717 &BD+56\,\,502      &   B0.7\,IIIe	 &$126\pm40$ &$26000\pm2000$      &       $3.8\pm0.2$   \\
662 &HD\,14052   &       B1\,Ib-II       &$100\pm20$   &$26500\pm700$        &       $3.41\pm0.08$  \\
339 &HD\,13969   &       B1\,III &$70\pm20$    &$25000\pm1000$        &       $3.7\pm0.1$  \\
922 &BD+56\,\,515      &   B1\,III-IVe	 &$253\pm20$   &$15000\pm1000$      &       $2.3\pm0.2$  \\
1133&BD+56\,\,526      &   B1\,III &$257\pm14$	 &$21000\pm2000$      &       $2.9\pm0.2$   \\
992 &BD+56\,\,520      &   B1\,V &$192\pm40$	 &$17900\pm800$      &$3.1\pm0.1$ \\
1132&BD+56\,\,527      &   B1.5\,II      &$114\pm17$   &$25100\pm600$  &       $3.4\pm0.1$  \\
1116&BD+56\,\,525      &   B1.5\,III	 &$134\pm20$   &$23000\pm1000$  &	 $3.39\pm0.06$  \\
846 &BD+56\,\,509      &   B1.5\,III-IVe   &$184\pm30$   &$25000\pm1000$	 &	 $3.64\pm0.09$  \\
843 &BD+56\,\,510      &   B1.5\,IV	 &$156\pm80$   &$22000\pm2000$	 &	 $3.6\pm0.2$	 \\
980 &BD+56\,\,519      &   B2\,III-IV	 &$64\pm20$	 &$21000\pm1000$	 &	 $3.6\pm0.1$  \\
\hline
\hline
\end{tabular}
\tablefoot{
Stellar parameters based on Astro+. Characteristics of individual stars:\\
\tablefoottext{a}{V757 Per; HD 14053}
}
\end{table*}

Astro+ relies on a grid of {\sc fastwind} models containing only H and He lines, and uses the strengths and shapes of selected diagnostic lines to determine physical parameters. The resulting parameters for the subset analysed are listed in Table~\ref{tab:param869ASTRO}. In Fig.~\ref{fig:comparisonTg}, we compare the results obtained with the two methods. Overall, the agreement is good. Fast rotators tend to drive both codes towards low-gravity solutions. For star 846 (BD +56 509), the global fit yields a low gravity typical of a supergiant, and consequently a very cool $T_{{\rm eff}}$ for the spectral type. Contrarily, Astro+ gives parameters fully consistent with the spectral type. Conversely, for 922 (BD +56 515), the global fit gives parameters typical of the spectral type, while Astro+ yields very low $T_{\mathrm{eff}}$ and $\log\,g$. 
Both 846 and 922  are rapidly rotating stars, whose shallow, rotationally broadened lines make the $T_{\mathrm{eff}}$-$\log\,g$ determination strongly degenerate, yielding the broad, elongated $\chi^2$ minima described in \citet[][their fig.~7]{Rubke2026}. In such cases, either procedure can settle on a spuriously cool, low-gravity solution. This is aggravated by the fact that the Astro+ FASTWIND grid currently includes only H and He lines, so that for fast rotators the Si ionisation balance -- which helps break the $T_{\mathrm{eff}}$-$\log\,g$ degeneracy in the global fit -- is not available. In both cases the spectral type provides the more reliable anchor. We therefore regard these two objects as illustrating the limits of automated parameter determination for fast rotators, and exclude them when quantifying the agreement between the two methods. Leaving them out, the two parameter sets agree with mean differences of $\sim600$\:K and $-0.07$\:dex and dispersions of
$\sim2000$\:K and $\sim0.2$\:dex, comparable to the individual uncertainties (Fig.~\ref{fig:comparisonTg}).

\section{Discussion}
\label{sec:discussion}

After obtaining astrophysical parameters for most stars in our sample, we built both the  conventional Hertzsprung-Russell (HRD) and the spectroscopic HRD (sHRD). The sHRDs \citep[e.g.][]{langerku2014,castro2014} for the two clusters are shown separately in Fig.~\ref{fig:brottdiags}. For the HRD, we restricted the sample to stars with reliable photoelectric photometry, with consistent measurements among several references, and excluded Be stars (which are variable and present intrinsic reddening due to their circumstellar disks). As the number of stars is much smaller, and the chances of having two stars with exactly the same parameters much lower, we display stars from the two clusters in the same plot (Figure~\ref{fig:brott_photo}). Stellar luminosities were calculated using the intrinsic colour to temperature relation of \citet{Flower1996}, adopting a standard reddening law to estimate the extinction and applying the bolometric correction scale of \citet{Flower1996} at the cluster distance. From the two diagrams, it is clear that both clusters contain an essentially identical population of stars -- although the sample is much smaller for NGC~884 -- except for the very luminous BSGs found in NGC~869. In what follows, we interpret the observed properties of the two populations in the context of recent stellar evolutionary models.

\subsection{Upper main sequence population}

\begin{figure*}
\resizebox{\columnwidth}{!}{\includegraphics[angle=0]{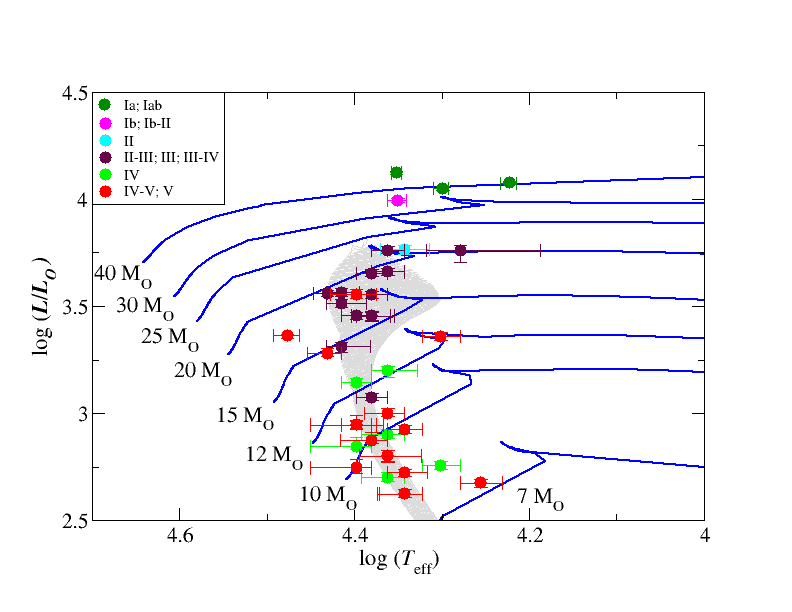}}
\resizebox{\columnwidth}{!}{\includegraphics[angle=0]{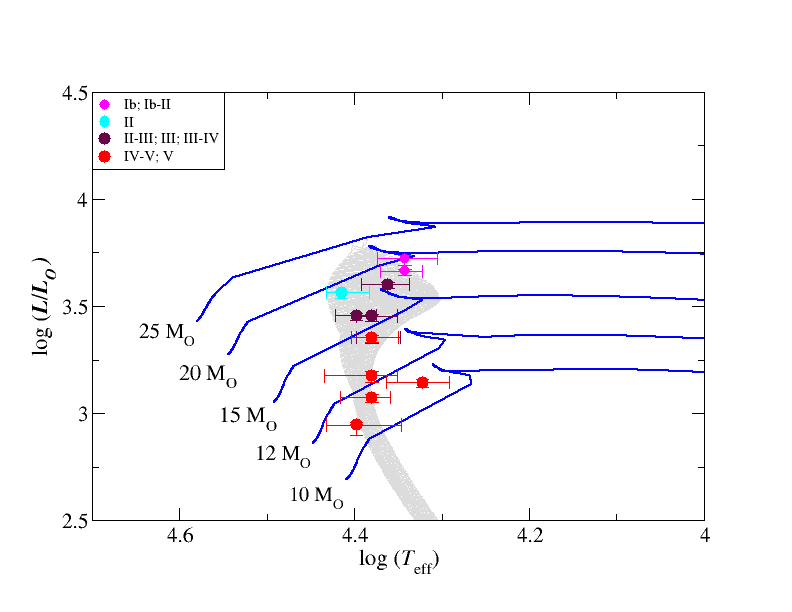}}
\caption{ Spectroscopic HRD \citep{langerku2014} for stars in the Double Cluster. Stars classified as luminosity class IV-V and V are shown in red; those in luminosity class IV are in light green; and giants with classes II-III, III, and III-IV are in brown. Objects classified as class II are shown as cyan circles. The supergiants stars with luminosity classes Ib and Ib-II are marked with magenta circles and those with luminosity classes Ia and Iab are marked with green circles. 
The continuous lines are evolutionary tracks with solar metallicity and $v_{\mathrm{rot}}\approx 170\:\mathrm{km}\,\mathrm{s}^{-1}$ from \citet{brott2011}, labelled with the stellar mass. The shaded area indicates the locus of parameters compatible with 12~Ma isochrones for a wide range of initial rotational velocities, after \citet{brott2011}.  
Left: Stars in NGC~869. Right: Stars in NGC~884
\label{fig:brottdiags} }
\end{figure*}

\begin{figure}
\resizebox{\columnwidth}{!}
{\includegraphics[angle=0]{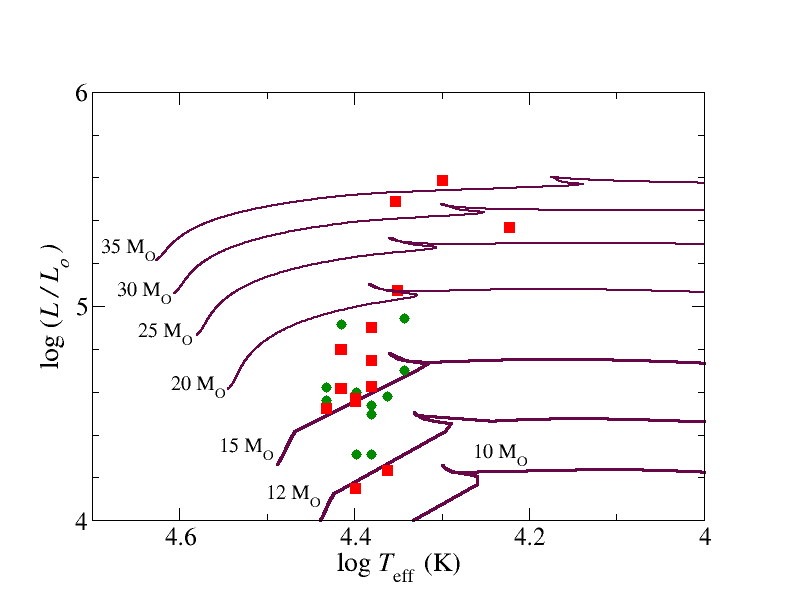}
}
\caption{Standard HRD for the subset of objects that have reliable $UBV$ photometry in the literature but with Be stars excluded, because of their variability. The luminosity was calculated by assuming the distance from \cite{Hunt2023}, a standard extinction law,  and the bolometric correction from \citet{Flower1996}. Stars considered members of NGC~869 are marked with red squares, while green circles are members of NGC~884.}
\label{fig:brott_photo}
\end{figure}

The Double Cluster lies in the nuclear region of Per~OB1. As shown in Table~\ref{tab:parametersHunt}, the astrometric parameters of NGC~869 and NGC~884 are essentially indistinguishable. Consequently, assigning a given star to one or the other necessarily depends on assumptions and the algorithm adopted for membership determination. Moreover, many of the stars with \citet{Oosterhoff1937} numbers lie very far away from the cluster centres, and their assignment to a cluster or the broad association is also somewhat arbitrary. Nevertheless, this ambiguity is of limited importance for the present analysis, because most recent studies have shown that the differences in age between the populations of the two clusters (and the surrounding association) are very small, and can be considered negligible in the central region studied here \citep[e.g.][]{Slesnick2002,Currie2010,Marsh2012}. We can therefore treat all the stars as members of a single stellar population, although we have plotted the two clusters separately for clarity in Fig.~\ref{fig:brottdiags}. In this figure, stars are separated by luminosity class, as this is a quick way of estimating the evolutionary stage of a given star. For reference, in Fig.~\ref{fig:rotators} we present the same sHR diagrams, but this time displaying the approximate rotational velocity of the stars and marking separately Be stars and the blue straggler. The information in the two figures must be considered in conjunction. For example, 2227 is marked as a star with moderate rotation, as its observed rotational velocity ($v_{\mathrm{rot}} = 160\:\mathrm{km}\,\mathrm{s}^{-1}$) falls in this range. But this is a very high rotational speed for a supergiant, as reflected in its spectral type B1.5\,Ibn.

\subsection{The meaning of the turn-off}

In photometric colour-magnitude diagrams, the cluster turn-off is identified with the bluest star(s) on the stellar sequence. This is not very strongly reflected in the CMDs displayed in Fig.~\ref{fig:Gaiadiagrams}, because the sensitivity of ($BP-RP$) to temperature is even weaker than in the traditional Johnson ($B-V$) \citep[cf.][]{massey1995}. In spectroscopic diagrams, the equivalent would be the star with the highest effective temperature. For old clusters, the turn-off is generally assumed to represent the point where stars reach the end of the main sequence and start moving towards the giant branch. In young clusters, core-hydrogen burning extends well beyond the observed turn-off point. Traditionally, B-type stars just above the turn-off are given luminosity class IV. In NGC~869, all these definitions broadly agree: There are a few B1\,V dwarfs (the hottest stars in the cluster), while luminosity class IV stars have spectral types B1.5 or B2. In NGC~884 there are no luminosity class IV stars, and B1\,V objects occupy the same region of the sHRD, between the 10 and $12\:\mathrm{M}_{\sun}$ tracks, which is in agreement with the expected mass for these objects \citep[for example, B1\,V stars typically have $11\:\mathrm{M}_{\sun}$ according to][calibrated with eclipsing binaries]{harmanec1988}. The spectra of these stars are displayed in Figures~\ref{fig:to_h} and~\ref{fig:to_chi}. 

Well above this turn-off, there is a significant population of giants, most of spectral types B1 and B1.5. These objects concentrate around the $15\:\mathrm{M}_{\sun}$ track, most of them close to the blue hook, which marks the end of the core hydrogen phase in these models. In fact, if we use the standard-rotation ($\Omega_{\mathrm{ini}}=0.4$) Geneva isochrones \citet{ekstrom2012}, most giants fall within the Hertzsprung gap, leading us to prefer the \citet{brott2011} models \citep[see][for a more general discussion]{castro2014}. Recent work \citep[e.g.][]{deBurgos2025} suggests that the TAMS lies at even lower temperatures than indicated by the \citet{brott2011} models\footnote{The recent work by \citet{sciarini2026} represents a first attempt to reproduce this observational TAMS in evolutionary grids.}. The concentration of Be stars in this region of the sHRD is consistent with the idea that they are most frequent close to the end of main sequence.

At higher luminosities, we find a few stars of luminosity classes II or Ib, which fall close to the blue hook of the $20\:\mathrm{M}_{\sun}$ track. The spectra of these stars are displayed in Fig.~\ref{fig:luminous}. Finally, there are two very luminous Ia supergiants in the core of NGC~869, and two other luminous $\sim$B1 supergiants close to its tidal radius, which fall above the $30\:\mathrm{M}_{\sun}$ track.  

\subsection{The presence of blue stragglers}
\label{sec:bs}

In photometric diagrams, blue stragglers are luminous objects located to the left of the cluster isochrone \citep[e.g.][for NGC~869]{Marco2001}. Among our sample, there is one obvious spectroscopic blue straggler (SBS), the B0\,V star 245 (BD+56 481), which also falls clearly to the left of the sequence in Fig.~\ref{fig:brottdiags} (left panel)\footnote{Oo~245 is not included in Fig.~\ref{fig:brott_photo}, as it fails the photometry-quality criteria. It is clearly identified in Fig.~\ref{fig:rotators} (left panel)}. This star is far away from the cluster core (see Fig.~\ref{fig:bigmap}) and its inconsistent proper motions suggest that it belongs to the wide association. Its shallow lines may indicate that it is a spectroscopic binary.

It could also be argued that the B0.7 giants are also SBS, since their spectral type is earlier than B1, which defines the turn-off. Nevertheless, when we consider the spectroscopic diagrams, they do not stand out as blue stragglers: their atmospheric parameters are not significantly different from those of the B1 giants. The most interesting case is Star 717 (BD +56 502), classified as B0.7\,IIIe. Although this star lies well within the tidal radius of NGC~869, its proper motions are atypical of the association, suggesting a recent dynamical perturbation. According to \citet{strom2005}, this star displays important variations in radial velocity, and it may be a binary.

In a spectroscopic diagram, the definition of a blue straggler depends on the adoption of an age for the population displayed. As a reference, in Fig.~\ref{fig:brottdiags}, we have shown the region delineated by the 12~Ma isochrones with a wide range of initial rotational velocities \citep[][]{brott2011}. Older isochrones place most of the B1\,V stars to their blue side. Although 12\:Ma is slightly younger than previous determinations, which favoured ages 13\,--\,15~Ma \citep[e.g.][]{Marco2001,Slesnick2002,Currie2010} and clearly inconsistent with the 18\:Ma favoured by \citet{Hunt2023}, this age is referred to a different set of stellar models and may not be directly comparable. In Fig.~\ref{fig:brottdiags}, all stars fainter than $G=8.0$ are compatible with this age, except for 245 and a few fast rotators (easily identifiable in Fig.~\ref{fig:rotators}) for which we find lower temperatures than typical of their spectral types (see Section~\ref{sec:params} for a discussion of the reasons. The four bright supergiants, in contrast, are completely incompatible with this isochrone. Their stellar parameters are similar to those of stars with the same spectral types in the younger cluster Westerlund~1 ($\tau\approx6$~Ma). In Westerlund~1, this B1\,--\,2 luminous supergiants follow the isochrone defined by the main-sequence stars \citep{Castellanos2026}.

\subsection{Single versus binary tracks}

\subsubsection{The supergiants}

The luminous supergiants near the core of NGC~869 (HD~14134 and HD~14143) and the nearby B1\,Iab HD~13854\footnote{As discussed by \citet{Negueruela2024}, HD~13854 is close to the limit with a Ia supergiant.} are very well-studied objects, whose stellar parameters have been derived by different authors based on diverse datasets \citep[e.g.][]{crowther2006,searle2008,deBurgos2023}. In all cases, their parameters place them in the HRD above the $25\:\mathrm{M}_{\sun}$ track, and, in the case of HD~14143 and HD~13854, well above the $30\:\mathrm{M}_{\sun}$ track (see e.g.  Fig.~\ref{fig:brott_photo}). In the sHRD (Fig.~\ref{fig:brottdiags}; left), their locations are broadly similar, although HD~14134 seems now much more massive (such discrepancies are normally taken as indication of a non-standard evolutionary history). In any event, all three objects are much more massive than any other star in our sample by such a large amount that any definition of the TAMS that we may want to adopt does not allow single stellar evolution tracks to reach such high luminosities. Therefore, these BSGs can only be interpreted as the product of binary evolution. Both HD~14134 and HD~14143 are considered members of NGC~869, while HD~13854, though formally outside its tidal radius, has astrometric parameters fully compatible with the cluster population.

As in NGC~663, these BSGs show no evidence of having been dynamically perturbed, which strongly favours a merger origin, since a mass-gainer scenario through any other process would require a former primary so massive that it should have exploded as a supernova. In addition, the very high inferred masses of these objects, even allowing for the possibility that they are overluminous for their current masses, require the merger of two stars with masses above the cluster turn-off. Moreover, \citet{deBurgos2023} derive significant He enrichment for both HD~14134 and HD~13854. Such enrichment is considered by \citet{menon2024} a telltale sign of a merger. 

The case of HD~13841 is less evident. Although its $\varpi$ and pmRA are typical of cluster members, its pmDec is slightly outside the 3$\sigma$ limit, and the star lies beyond the cluster tidal radius. More interestingly, the parameters derived for HD~13841 are similar within uncertainties to those of HD~13854, even though the latter is over a magnitude brighter. Although HD~13841 has not been studied as much as the other BSGs, older works confirm this similarity in parameters \citep[e.g.][]{Kudritzki1999}. The enormous discrepancy between the position in the sHRD (together with the Ia supergiants) and the HRD (together with the other Ib supergiants) suggests that this star has not followed a standard evolutionary path either. However, its parameters and diverse proper motion allow for scenarios other than a merger \citep[see][for possible evolutionary scenarios]{Wang2020}.

Given the strong evidence of a binary origin of the luminous supergiants, we must next consider the evolutionary status of the stars close to the $20\:\mathrm{M}_{\sun}$ track. This is a population of B1\,--\,B1.5 objects of luminosity classes Ib and II. Interestingly, all of them have kinematics compatible with the clusters, with the brightest stars in NGC~884 (2227 and 2296) being 1$\sigma$ members. For these objects, however, the possibility of a merger appears considerably less compelling. They lie at the top of the continuous cluster sequence, while the luminous supergiants are clearly detached from the rest. They are also compatible with the same isochrone as the rest of the stars. Finally, \citet{deBurgos2024} do not find evidence of He enhancement in any of them.

\subsubsection{The Be stars}
\label{sec:be_evol}

At slightly lower masses, we find a large number of giants, a significant fraction of which are Be stars. Some of them are very luminous Be stars. If they have followed single-star evolutionary tracks, they started their lives as O9\,--\,B0 dwarfs \citep{Martins2017}. Be stars of luminosity class II are exceedingly rare \citep{negueruela2004}, and therefore, BD+56 473 must be one of the most luminous Be stars known. It was originally classified B1\,II: by \citet{Morgan1955}, and here we put it at B1\,II--III. Another very luminous Be star is BD +56 511, B1.5\,II--III. Nevertheless, Be stars in the area cover a wide range of magnitudes.

Table~\ref{tab:bestars} lists all the confirmed Be stars within the tidal radii of NGC~869 and NGC~884 \citep[as derived by][]{Hunt2023}, as well as those immediately outside them. More than half of the Be stars in the area have $G$ magnitudes between 9 and 10, corresponding to stars above the turn-off. The presence of Be stars above the turn-off may be explained by angular momentum transport from the core to the envelope during single-star evolution  \citep[e.g.][]{granada2013,hastings2020}, although this mechanism is currently not believed to be the main channel to form Be stars. Mass and angular momentum transfer in a binary are believed to dominate Be star formation \citep[e.g.][and references therein]{wang2023}.

One of the strongest arguments in favour of a binary origin for Be stars is the presence of Be stars below the turn-off in young clusters \citep{wang2023}, although there is some evidence that late-B stars may be born with high initial rotational speeds \citep{Rivi2026}. In Table~\ref{tab:bestars}, objects below the horizontal line are with certainty below the turn-off, with spectral types B2\,V and later. There are nine such stars, a small, but not insignificant number. Two of them (1058 and 1912) have magnitudes typical of late-B stars, but all the others are more than one magnitude brighter, suggesting they are mid-B stars. Interestingly, all nine have kinematics consistent with the clusters.

Over the past decades, many population synthesis studies have tried to explain the formation of Be stars through mass accretion in binary systems. Most models developed seek to reproduce the observed population of Be/X-ray binaries. As such, they are strongly constrained, because the binary must remain bound after the supernova explosion that forms the neutron star. This favours relatively conservative mass transfer, because a more massive Be star is more likely to retain its companion. It also favours a second phase of mass transfer, after the initial primary has evolved into a He star, because a less massive core reduces the amount of mass lost in the explosion and may increase the probability of a relatively symmetric explosion. Of course, this implies that there will be many other systems unbound by the supernova explosion, leading to a population of runaway stars, many of which should be Be stars as a consequence of their previous mass accretion.

These constraints are loosened when the binary companion to the Be star is a He star, which has not undergone core collapse yet, or is not massive enough to explode. Nevertheless, population-synthesis calculations aimed at reproducing the properties of observed systems continue to favour moderately conservative case B evolution \citep[e.g.][]{schootemeijer2018,Shao2021,Lechien2025}. The observational biases are different, though, because in this case a more massive Be star disfavours the detection of the He star companion. According to \citet{Shao2021}, only if accretion is assumed to stop as soon as the accretor reaches its critical rotational velocity, can case A evolution contribute as much as case B to the formation of Be + He star systems. However, the overall properties of known systems are better reproduced by moderately conservative case B evolution.

Given the masses of the Be giants in our sample, and the fact that they are close to the TAMS, case B evolution must have happened in binaries with a primary initially more massive than $\approx15\:\mathrm{M}_{\sun}$. Such primaries are expected to undergo core-collapse supernovae even after heavy mass loss. If this evolutionary channel dominates, a substantial fraction of Be stars above the turn-off should have anomalous kinematics due to the supernova disruption of their former binary systems. Contrarily, the majority of the Be stars in the clusters have proper motions compatible at the 3$\sigma$ level with the cluster averages. Interestingly, however, five of the seven brightest Be stars have incompatible kinematics. This could point to a binary origin for them. 

Nevertheless, the proper motions of these stars do not point to an ejection from the densest parts of the clusters. Of all the Be stars with anomalous proper motions, only 2284 has a motion that might be compatible with an ejection from the central areas of the respective clusters. For all others, the direction of motion is not adequate; for example, 2402 is moving towards NGC~884 from the north, while 847, the brightest Be star, seems to have an eastwards-directed motion that passes South of the core of NGC~869. As mentioned above, 717 lies in the central region of the cluster and presents perturbed kinematics. But many of the giants that do not show the Be phenomenon also show perturbed kinematics. Among them, 339 (HD~13969) may be an ejection from NGC~869, but its proper motions are too unreliable (RUWE$\approx10$) to ascertain this from DR3 data.

If the formation of massive Be stars through binary mass transfer is commonly followed by the supernova explosion of the former primary, one of the expected signatures should be a population of Be stars dynamically ejected from the cluster cores. Inspection of the Be stars in the catalogue of \citet{Jaschek1982} out to $\approx 1\degr$ from the cluster cores shows no compelling evidence of this. None of them has proper motions that provide convincing support for an ejection scenario, with all being compatible with the dynamics of the association. Additionally, as in NGC~663, none of the Be stars in the Double Cluster is a blue straggler, with the possible exception of 717, discussed above. The distribution of spectral types among the Be stars is indistinguishable from that of non-emission stars -- except for 517, discussed in Section~\ref{sec:bes}, which may be later. Taken together, the evidence of case B mass transfer followed by a supernova explosion as the main channel to form the massive Be stars is, at best, weak.

For the Be stars around and below the turn-off, the kinematics do not support even moderate perturbations. Therefore, a strong contribution of case B evolution is difficult to reconcile with the observations, unless the mass donors have lost so much mass loss that the cores of stars initially with $M_{*} \gtrsim 15\:\mathrm{M}_{\sun}$ have become too light to produce a supernova explosion. It seems more natural to assume that these Be stars below the turn-off are preferentially formed via case A mass transfer in very close binaries.

\section{Conclusions}
\label{sec:conclus}

We have derived stellar parameters for almost 60 stars in the upper main sequence (B2\,V and earlier) in the region of the Double Cluster, h \& $\chi$~Persei, most of them in the cluster cores, by using an automated spectral-fitting procedure. We have compared the outcome of two independent automated analysis procedures and find broad agreement in the parameters derived. Parameters for fast rotators tend to be inconsistent with the observed spectral types, sometimes including unphysically low gravities that lead to low temperatures, as well. This behaviour, however, is not systematic with a given code failing for a spectrum but giving parameters consistent with expectations for a similar one.

A surprisingly large fraction (around 40\%) of the stars along the photometric sequence of NGC~869 have proper motions incompatible with the cluster average at the 2$\sigma$ level. This does not happen for NGC~884 (where the fraction of such stars is $\lesssim20$\%) or in any other young open cluster that we have examined. This suggests a strong dynamical perturbation in this field, for which we find no obvious explanation. In view of this, kinematic anomalies alone cannot be regarded as conclusive evidence for a binary origin of individual stars in this region, and interpretations based on their proper motions should be treated with caution. 

We have placed all stars in the (spectroscopic) HR diagram and analysed their locations. When our results are combined with those for the two most massive clusters in the Perseus arm, NGC~7419, which has an age similar to that of the Double Cluster \citep{marco2013}, and NGC~663, which is older and more evolved \citep{Marco2025}, we find that:
\begin{itemize}
    \item The bulk of the population in all the massive clusters is fully consistent with single-star evolution at a common age, particularly when a range of initial rotational velocities is considered. The number of convincing blue stragglers is very small: one in NGC~7419, perhaps one in NGC~869, none in NGC~884, and four in NGC~663.
    \item Both NGC~869 and NGC~663 contain a small population of BSGs that are much more luminous than any other star in the clusters. Their kinematics are fully consistent with cluster membership, and their stellar parameters are similar to those of stars with the same spectral types in significantly younger clusters. These properties strongly favour a merger origin.
    \item All the clusters younger than $\approx15$~Ma contain a population of low luminosity supergiants (with classes Ib or II) crowning the stellar sequence. These stars have typical cluster kinematics and may represent the most luminous single stars or, less likely, the mergers of stars with masses below the turn-off.
    \item In all four clusters, the spectral-type distribution of Be stars is essentially indistinguishable from that of the non-emission stars, which hints at common formation mechanisms. Only the Be/X-ray binary LS~I~$+61\degr$\,235 in NGC~663 and possibly 717 in NGC~869 are mild blue stragglers. Moderately conservative case B evolution would, on the other hand, be expected to produce at least some rejuvenated main sequence stars that are significantly more massive than any single star and, hence, present an earlier spectral type.
    \item Be stars below the turn-off, which cannot be produced via single-star evolution, have kinematics typical of cluster members, indicating that they have not been perturbed by the supernova explosion of a massive initial primary. This strongly favours case A mass transfer in binaries below the turn-off as the formation channel for these systems.
    \item We find no convincing population of Be stars ejected from either NGC~663, NGC~869, or NGC~884\footnote{Due to much higher reddening, the stars in NGC~7419 are too faint to be able to hold a similar conclusion.}, except for the Be/X-ray binary LS~I~$+61\degr$\,235. This is, again, difficult to reconcile with a scenario in which moderately conservative case B evolution is the dominant formation channel for massive Be stars.
\end{itemize}

The forthcoming Gaia DR4 release will probably improve the membership analysis for some of the stars with poor astrometric solutions. However, given the high quality of most proper motions in EDR3, it is highly unlikely to have any effect on our analysis.

\begin{acknowledgements}
Many of the observations used here were obtained as part of several poor-weather backup programmes running at the WHT. We thank the support astronomers for their efforts. We also thank the referee, Nikolay Britavskiy, for helpful suggestions to make the paper easier to follow. \\
This research is partially supported by the Spanish
Government Ministerio de Ciencia, Innovaci\'on y Universidades and Agencia Estatal de Investigación (MCIU/AEI/10.130 39/501 100 011 033/FEDER, UE) under grant PID2024-159329NB-C21/22. It is also supported by the Generalitat Valenciana under grant CIAICO/2024/186. NC acknowledges funding from the Deutsche Forschungsgemeinschaft (DFG) — CA 
2551/1-2.\\
This research has made use of the SIMBAD, Vizier and Aladin
services developed at the Centre de Données Astronomiques de Strasbourg (CDS), France, and the VO TOPCAT \citep{TOPCAT2005} tool. This work has made use of data from the European Space Agency (ESA) mission Gaia (https://www.cosmos.esa.int/gaia), processed by the Gaia Data Processing and Analysis Consortium (DPAC, https://www.cosmos.esa.int/
web/gaia/dpac/consortium). Funding for the DPAC has been provided by
national institutions, in particular the institutions participating in the Gaia Multi-lateral Agreement. 
\end{acknowledgements}

\begin{appendix}
\onecolumn
\section{Tables}
\begin{longtable}{rllccc}
\caption{Spectral types and stellar parameters for stars in NGC~869.}
\label{tab:param869}\\
\toprule
Oo &
Other &
Spectral &
 $v\,\sin\,i$ &
$T_{\mathrm{eff}}$ &
 $\log\,g$\\
Number  &
Name &
Type &
(km\,s$^{-1}$) &
(K) &
 (dex)\\
 \midrule
 \endfirsthead
 \caption[]{continued.}\\
 \toprule
 Oo &
Other &
Spectral &
 $v\,\sin\,i$ &
$T_{\mathrm{eff}}$ &
 $\log\,g$\\
 Number&
 Name&
Type &
(km\,s$^{-1}$) &
(K) &
 (dex)\\
 \midrule
 \endhead
 \midrule
\multicolumn{6}{l}{Continued on next page}
\endfoot
\bottomrule
\endlastfoot
245 &BD+56\,481     &		B0\,V\tablefootmark{a}  &140	&$30000^{+1100}_{-1000}$	&	$3.90^{+0.13}_{-0.10}$  \\
612 &BD+56\,498\tablefootmark{b}     &   B0.7\,II&80	 &$24000^{+1000}_{-1000}$&	 $3.22^{+0.10}_{-0.10}$  \\
260 &BD+56\,482      &   B0.7\,III	 &180	 &$26000^{+1000}_{-1200}$      &       $3.45^{+0.12}_{-0.10}$  \\
692 &BD+56\,501      &   B0.7\,III	 &200&  $27000^{+1000}_{-1000}$   &	 $3.52^{+0.10}_{-0.10}$  \\
717 &BD+56\,502      &   B0.7\,IIIe	 &120   &$26000^{+1000}_{-2000}$      &       $3.7^{+ 0.1}_{- 0.2}$   \\
839 &BD+56\,508      &   B1 III\,      &40     &$24000^{+1300}_{-1200}$        &       $3.42^{+0.11}_{-0.17}$  \\
662 &HD14052   &       B1\,Ib-II       &100   &$24000^{+1000}_{-1000}$        &       $3.22^{+0.10}_{-0.10}$  \\
49  &BD+56\,473      &   B1\,II-IIIe\tablefootmark{c}&&Strong emission        &       \\
847 &BD+56\,511      &   B1.5\,II-IIIe   &	100 &$24000^{+1000}_{-1000}$      &       $3.32^{+0.10}_{-0.13}$  \\
339 &HD~13969   &       B1\,III &100    &$26000^{+1000}_{-1700}$        &       $3.60^{+0.15}_{-0.10}$  \\
803 &BD+56\,505      &   B1\,III-IVnn	 &320	 &$24000^{+1000}_{-1400}$      &       $3.42^{+0.10}_{-0.14}$  \\
922 &BD+56\,515      &   B1\,III-IVe	 &280   &$24000^{+1100}_{-1700}$      &       $3.22^{+0.10}_{-0.10}$  \\
1268&HD~14162   &       B1\,III\tablefootmark{d}& 140   &$25000^{+1000}_{-1000}$        &       $3.49^{+0.10}_{-0.10}$  \\
1133&BD+56\,526      &   B1\,III &220	 &$23000^{+1000}_{-1000}$      &       $3.04^{+0.10}_{- 0.10}$   \\
1141&BD+56\,528      &   B1\,Vn  &	 &	No spectrum &				       \\
496 &BD+56\,488      &   B1\,V\tablefootmark{a}  & &Likely SB2     &      \\
566 &BD+56\,493      &   B1\,V\tablefootmark{d}  &330	 &$25000^{+1700}_{-1300}$      &       $3.39^{+0.13}_{-0.12}$  \\
992 &BD+56\,520      &   B1\,V&240 &$20000^{+1000}_{-1000}$      &$3.20^{+0.10}_{-0.10}$ \\
1078&BD+56\,524      &   B1\,IV-V      &200    &$27000^{+1400}_{-1000}$        &       $3.80^{+0.15}_{-0.10}$  \\
1132&BD+56\,527      &   B1.5\,II      &120   &$22000^{+1400}_{-        1400}$  &       $2.96^{+0.10}_{-0.10}$  \\
678 &LSV +57\,9      &	  B1.5\,IV	&80	&$23000^{+1000}_{-1800}$	&	$3.60^{+0.10}_{-0.22}$  \\
1116&BD+56\,525      &   B1.5\,III	 &140   &$23000^{+1000}_{-1000}$  &	 $3.14^{+0.10}_{-0.10}$  \\
309 &BD+56\,484      &   B1\,III-IVe	  & &Strong emission&  \\
846 &BD+56\,509      &   B1.5\,III-IVe   &220   &$19000^{+1700}_{-4000}$	 &	 $2.71^{+0.10}_{-0.19}$  \\
1261&BD+56\,534      &   B1.5\,IVe	 &   &Strong emission&  \\
843 &BD+56\,510      &   B1.5\,IV	 &140   &$25000^{+1000}_{-1000}$	 &	 $3.80^{+0.10}_{-0.10}$	 \\
777 &     &	  B1.5\,V &100    &$24000^{+2000}_{-1000}$	  &	  $4.00^{+0.22}_{-0.10}$	  \\
864 &BD+56\,513      &   B1.5\,V&220	 &	 $23000^{+1400}_{-1000}$ &	 $3.80^{+0.14}_{-0.10}$  \\
929 &BD+56\,516      &   B1.5\,V&180	 &	 $25000^{+1000}_{-1000}$ &	 $4.00^{+0.10}_{-0.10}$  \\
936 &BD+56\,517      &   B1.5\,V&40	 &	 $25000^{+1000}_{-1300}$ &	 $4.00^{+0.16}_{-0.10}$	 \\
1085&     &	  B1.5\,V &80	  &$25000^{+1000}_{-1400}$	  &	  $4.00^{+0.12}_{-0.10}$	  \\
980 &BD+56\,519      &   B2\,III-IV	 &100	 &$24000^{+1000}_{-1000}$	 &	 $3.80^{+0.10}_{-0.10}$  \\
1161&BD+56\,529      &   B2\,IVe &   &Strong emission &	  \\
1282&EM*GGA 156     &     B2\,IV-Ve	   &  &Strong emission  &	    \\
1364&BD+56\,538      &	  B2\,IV-V	  &280    &$25000^{+3000}_{-1000}$	  &	  $4.00^{+0.26}_{-0.10}$  \\
530 &BD+56\,491      &	  B2\,IV\tablefootmark{a}&& Possibly SB2   &  \\
622 &     &	  B2\,IV  &60&    $20000^{+1000}_{-1000}$ &	  $3.8^{+0.1}_{-0.1}$	  \\
670 &     &	  B2\,IV  &120&   $23000^{+1600}_{-1000}$ &	  $3.90^{+0.16}_{-0.13}$   \\
859 &     &	  B2\,IV  &220&   $20000^{+1000}_{-1000}$   &	  $3.8^{+0.1}_{-0.1}$	  \\
892 &     &	  B2\,IV  &60&    $23000^{+1600}_{-1000}$ &	  $4.10^{+0.11}_{-0.10}$    \\
978 &BD+56\,518      &	  B2\,IV  &40&    $25000^{+3000}_{-1000}$ &	  $4.10^{+0.26}_{-0.10}$    \\
879 &     &	  B2\,IV-V	  &100&   $22000^{+1600}_{-1000}$ &	  $4.1^{+0.1}_{-0.1}$	  \\
475 &     &	  B2\,V\tablefootmark{a}  &120&   $18000^{+1000}_{-1000}$   &	  $3.90^{+ 0.16}_{-0.10}$   \\
491 &     &	  B2\,Vn  &260&   $18000^{+1000}_{-1000}$	  &	  $3.7^{+0.1}_{-0.1}$	  \\
635 &BD+56\,496      & B1.5\,--\,B2\,V	 &300&   $22000^{+1000}_{-1000}$ &	 $3.80^{+ 0.12}_{-0.10}$   \\
977 &     &	    B2\,V &		  &	Not analysed  &				  \\
991 &     &	    B2\,V &120    &$22000^{+1500}_{-1000}$	  &$4.10^{+0.11}_{-0.10}$   \\
1004&     &	  B2\,V   &140    &$23000^{+1700}_{-2000}$	  &	  $4.00^{+0.17}_{-0.14}$	  \\
1109&	   &	    B2\,V  &240    &$23000^{+1700}_{-2000}$	   &	   $4.00^{+0.16}_{-0.18}$     \\
1041&	   &	   B2\,V   &200    &$22000^{+1000}_{-1300}$	   &	   $4.0^{+0.1}_{-0.1}$	   \\
1080&V617~Per	   &	   B2\,V   &180&   $25000^{+3000}_{-1000}$ &	   $4.2^{+0.2}_{-0.2}$     \\
1438&     &	  B2\,Ve  &    &	Strong emission  &	   \\
976 &     &	  B2.5\,V &	 &Poor spectrum&   \\
985 &     &	  B2.5\,V &80	  &$18000^{+1000}_{-1000}$	  &	  $4.00^{+0.12}_{-0.10}$	  \\
517 &BD+56\,489      &	  B3\,III shell   &  &	Shell spectrum &\\
649&     &	  B3\,IV\tablefootmark{e} &  & Late-type   &\\
250 &     &	  B3\,V   &   & Late-type   &   \\
880 &     &	  B3\,V   &	  & Late-type   &	  \\
893 &     &	  B3\,V   &		  & Late-type   &\\
907&&B4\,V&  & Late-type   &\\
930&&B4\,V   &   & Late-type   &    \\
1020&&B7\,V   &  & Late-type   &  \\
264&&B8\,V\tablefootmark{e}   &   & Late-type   &  \\
\end{longtable}
\tablefoot{
Stellar parameters based on the global fit method. Characteristics of individual stars:\\
\tablefoottext{a}{likely binary};
\tablefoottext{b}{V757 Per; HD 14053};
\tablefoottext{c}{extreme Be};
\tablefoottext{d}{known Be, but not obvious in blue spectrum};
\tablefoottext{e}{non-member}
}
\begin{table}[h!]
\caption{Mean proper motion and parallax for NGC~869 and NGC~884.}
\label{tab:parametersHunt}
\centering
\begin{tabular}{ccccc}
\hline\hline
Cluster &
Members &
Plx &
 pmRA&
pmDEC \\
 &
 &
(mas) &
(mas\,·\,year$^{-1}$) &
(mas\,·\,year$^{-1}$) \\
 \noalign{\smallskip}
\hline
\hline
\noalign{\smallskip}
NGC~869 &403	     &    $0.40358860\pm0.01909736$ &      $-0.65444107\pm0.07848563$     &       $-1.14328379\pm0.07854335$     \\
NGC~884&287    &      $0.40240438\pm0.01894602$	     &      $-0.61389632\pm0.07875767$    &       $-1.14234992\pm0.06505492$      \\
 \noalign{\smallskip}
\hline
\end{tabular}
\tablefoot{The values are taken from \citet{Hunt2023}}
\end{table}
\setlength{\tabcolsep}{3pt}
\begin{longtable}{llccrrrr}
\caption{Astrometric parameters for stars in NGC~869 from DR3.}
\label{tab:gaia869}\\
\toprule
Oo &
Other &
 RA(J2000) &
 DEC (J2000) &
 G&
Plx &
 pmRA&
 pmDEC\\
Number  &
Name &
($^\circ$) &
($^\circ$)&
(mag) &
 (mas)&
 (mas/yr)
 &(mas/yr)\\
 \midrule
 \endfirsthead
 \caption[]{continued.}\\
 \toprule
Oo &
Other &
 RA (J2000)&
 DEC (J2000)&
 G&
Plx &
 pmRA&
 pmDEC\\
Number  &
Name &
($^\circ$) &
($^\circ$) &
(mag) &
 (mas)
 &(mas/yr)&
(mas/yr) \\
 \midrule
 \endhead
 \midrule
\multicolumn{8}{l}{Continued on next page}
\endfoot
\bottomrule
\endlastfoot
 245\tablefootmark{c,d}  &  BD+56\,481	&   34.396323348  &  57.085671502  & 9.84  &   $0.37\pm0.02$ &  $-0.93\pm0.01$ &     $-0.69\pm0.02$\\
 612\tablefootmark{a}  &  BD+56\,498&   34.596034468  &  57.010196169  & 8.39  &   $0.35\pm0.09$ &  $-0.55\pm0.07$ &	$-1.39\pm0.09$\\
 260\tablefootmark{c,d}  & BD+56\,482	&   34.413044330  &  57.199328601  & 9.29  &   $0.42\pm0.02$ &  $-0.90\pm0.01$ &     $-1.34\pm0.02$\\
 692\tablefootmark{b,d}  & BD+56\,501	&   34.624301477  &  57.150873081  & 9.33  &   $0.45\pm0.02$ &  $-0.65\pm0.01$ &     $-1.36\pm0.02$\\
 717\tablefootmark{a}  &  BD+56\,502	&   34.636577171  &  57.211019677  & 9.23  &   $0.40\pm0.03$ &  $ 0.69\pm0.02$  &    $-0.83\pm0.03$\\
 839\tablefootmark{c,d}  &  BD+56\,508	&   34.700070848  &  57.285518638  & 9.28  &   $0.41\pm0.02$ &  $-0.23\pm0.01$ &     $-1.42\pm0.02$\\
 662\tablefootmark{d}  & HD~14052 		&   34.617315811  &  57.208368208  & 8.11  &   $0.42\pm0.02$ &  $-0.61\pm0.02$ &     $-1.30\pm0.02$\\
 49   & BD+56\,473	&   34.240454402  &  57.130197500  & 8.95  &   $0.41\pm0.02$ &  $-0.67\pm0.01$ &     $-1.22\pm0.02$\\
 847\tablefootmark{c,d}  &   BD+56\,511	&   34.699737629  &  57.067270455  & 9.02  &   $0.39\pm0.02$ &  $-0.14\pm0.01$ &   $-1.01\pm0.02$\\
 339\tablefootmark{a} &   HD~13969	&   34.457686248  &  57.090427564  & 8.79  &  $-0.13\pm0.16$ &  $-0.36\pm0.12$  &  $-2.44\pm0.16$ \\
 803\tablefootmark{d}   &  BD+56\,505	&   34.678214445  &  57.072637005  & 10.44 &   $0.43\pm0.02$ &  $-0.72\pm0.01$ &   $-1.36\pm0.01$\\
 922\tablefootmark{b,c}  &  BD+56\,515	&   34.724414981  &  57.139504847  & 9.47  &   $0.40\pm0.01$ &  $-0.78\pm0.01$ &   $-1.20\pm0.01$\\
 1268 &  HD~14162		&   34.864440283  &  57.138231220  & 9.29  &   $0.40\pm0.01$ &  $-0.59\pm0.01$ &   $-1.09\pm0.01$\\
 1133\tablefootmark{d} &  BD+56\,526	&   34.792958213  &  57.127102703  & 8.91  &   $0.42\pm0.02$ &  $-0.64\pm0.02$ &   $-0.93\pm0.02$\\
 1141\tablefootmark{d} & BD+56\,528	&   34.794499456  &  57.024953675  & 9.78  &   $0.43\pm0.02$ &  $-0.62\pm0.01$ &   $-1.39\pm0.01$\\
 496\tablefootmark{a}  &  BD+56\,488	&   34.556324512  &  57.358570848  & 9.98  &  -$0.38\pm0.16$ &  $-0.89\pm0.14$  &  $-1.26\pm0.18$\\
 566\tablefootmark{b}  &    BD+56\,493	&   34.576472569  &  56.850614595  & 9.62  &   $0.35\pm0.02$ &  $-0.62\pm0.01$ &   $-1.24\pm0.02$\\
 992\tablefootmark{a}  &    BD+56\,520	&   34.750863931  &  57.145639587  & 9.92  &   $0.54\pm0.03$ &  $-0.26\pm0.02$  &  $-1.11\pm0.02$\\
 1078 &   BD+56\,524	&   34.776782308  &  57.126077278  & 9.73  &   $0.42\pm0.02$ &  $-0.74\pm0.01$ &   $-1.18\pm0.02$\\
 1132\tablefootmark{d} &   BD+56\,527	&   34.793527766  &  57.130578514  & 8.36  &   $0.41\pm0.02$ &  $-0.69\pm0.02$ &   $-0.95\pm0.02$\\
 678\tablefootmark{c}  &    LSV\,+57\,9	&   34.624565818  &  57.366509934  & 10.24 &   $0.39\pm0.02$ &  $-0.78\pm0.02$ &   $-1.17\pm0.02$\\
 1116 &    BD+56\,525	&   34.788511205  &  57.135087380  & 9.19  &   $0.40\pm0.02$ &  $-0.70\pm0.01$ &   $-1.09\pm0.01$\\
 309\tablefootmark{c.d}   &   BD+56\,484	&   34.436066953  &  56.899997098  & 9.36  &   $0.42\pm0.02$ &  $-1.11\pm0.01$ &   $-1.36\pm0.01$\\
 846  &   BD+56\,509	&   34.701545908  &  57.240252253  & 9.87  &   $0.43\pm0.01$ &  $-0.68\pm0.01$ &   $-1.08\pm0.01$\\
 1261\tablefootmark{e} &   BD+56\,534	&   34.860486337  &  57.078364118  & 9.35  &   $0.45\pm0.01$ &  $-1.16\pm0.01$ &   $-1.69\pm0.01$\\
 843\tablefootmark{d}   &   BD+56\,510	&   34.699052024  &  57.135202342  & 9.27  &   $0.39\pm0.02$ &  $-0.68\pm0.01$ &   $-0.97\pm0.02$\\
 777\tablefootmark{c,d}  &  	        &   34.668884068  &  57.297130497  & 10.85 &   $0.45\pm0.02$ &  $-0.75\pm0.01$ &   $-1.19\pm0.02$\\
 864  &  BD+56\,513	&   34.706168948  &  57.097208651  & 9.91  &   $0.43\pm0.02$ &  $-0.60\pm0.01$ &   $-1.17\pm0.02$\\
 929\tablefootmark{c,d}   &  		&   34.726310634  &  57.158201376  & 10.27 &   $0.43\pm0.02$ &  $-0.76\pm0.01$ &   $-1.27\pm0.02$\\
 936\tablefootmark{b,c}  &   BD+56\,517	&   34.732189654  &  57.151835438  & 10.36 &   $0.45\pm0.02$ &  $-0.85\pm0.01$ &   $-1.11\pm0.02$\\
 1085 &  	        &   34.779029707  &  57.151235025  & 10.38 &   $0.41\pm0.02$ &  $-0.55\pm0.02$ &   $-1.00\pm0.02$\\
 980\tablefootmark{b}  &   BD+56\,519	&   34.745726138  &  57.145083258   & 9.66  &  $0.45\pm0.01$ &  $-0.73\pm0.01$ &   $-1.14\pm0.01$\\
 1161\tablefootmark{a} &   BD+56\,529	&   34.806969273  &  57.128764927   & 10.00    &  $0.13\pm0.12$ &  $-0.46\pm0.09$ &   $-1.08\pm0.11$\\
 1282 &   EM*GGA\,156	&   34.870089869  &  57.117923915   & 10.94 &  $0.41\pm0.02$ &  $-0.58\pm0.01$ &   $-1.10\pm0.02$\\
 1364 &  BD+56\,538	&   34.916954393  &  57.091987175   & 9.86  &  $0.43\pm0.02$ &  $-0.66\pm0.01$ &   $-1.10\pm0.01$\\
 530  &  BD+56\,491	&   34.571119362  &  57.208628123   & 10.17 &  $0.43\pm0.02$ &  $-0.71\pm0.01$ &   $-1.24\pm0.02$\\
 622  &  		&   34.603943574  &  57.167880387  & 10.82 &   $0.40\pm0.02$ &  $-0.68\pm0.02$ &   $-1.04\pm0.02$\\
 670  &  		&   34.618883365  &  57.195702637  & 10.86 &   $0.42\pm0.02$ &  $-0.73\pm0.02$ &   $-1.21\pm0.02$\\
 859\tablefootmark{c,d}  &  		&   34.706427917  &  57.121913143  & 10.70  &   $0.39\pm0.02$ &  $-0.51\pm0.01$ &   $-1.41\pm0.02$\\
 892\tablefootmark{c}  &  		&   34.713390345  &  57.142451191  & 11.17 &   $0.37\pm0.02$ &  $-0.54\pm0.02$ &   $-1.23\pm0.02$\\
 978  &  BD+56\,518	&   34.745822928  &  57.157392314  & 10.59 &   $0.42\pm0.02$ &  $-0.66\pm0.01$ &   $-1.20\pm0.02$\\
 879\tablefootmark{d}  &  		&   34.710949911  &  57.144783946  & 11.53 &   $0.41\pm0.02$ &  $-0.59\pm0.01$ &   $-1.27\pm0.02$\\
 475\tablefootmark{c,d}  &  		&   34.544811880  &  57.229287003  & 11.29 &   $0.40\pm0.02$ &  $-1.04\pm0.02$ &   $-1.44\pm0.02$\\
 491\tablefootmark{c,d}  &  		&   34.550074242  &  57.161243885  & 11.01 &   $0.39\pm0.02$ &  $-0.55\pm0.01$ &   $-1.41\pm0.02$\\
 635\tablefootmark{a}  &  BD+56\,496	&   34.610598511  &  57.380307946  & 11.05 &   $0.39\pm0.07$ &  $-1.05\pm0.06$ &   $-1.56\pm0.08$\\
 977\tablefootmark{e}  &  		&   34.744249016  &  57.136588205  & 11.03 &   $0.47\pm0.02$ &  $-0.28\pm0.02$ &   $-0.89\pm0.02$\\
 991\tablefootmark{d}  &  		&   34.750960070  &  57.139631826  & 11.36 &   $0.41\pm0.02$ &  $-0.75\pm0.02$ &   $-1.29\pm0.02$\\
 1004\tablefootmark{e} &  		&   34.753888501  &  57.150776022  & 10.80  &   $0.43\pm0.02$ &  $-0.70\pm0.01$ &   $-1.18\pm0.02$\\
 1109\tablefootmark{e} &  		&   34.786901393  &  57.125629986  & 10.90  &   $0.47\pm0.03$ &  $-0.34\pm0.02$ &   $-1.34\pm0.03$\\
 1041\tablefootmark{b,c} &  		&   34.763449563  &  57.139110682  & 11.00    &   $0.43\pm0.02$ &  $-0.15\pm0.02$ &   $-1.16\pm0.02$\\
 1080 &    V617\,Per	&   34.777999390  &  57.147828377  & 11.08 &   $0.44\pm0.02$ &  $-0.70\pm0.02$ &   $-1.20\pm0.02$\\
 1438 &  		&   34.949340914  &  57.110979378  & 10.93 &   $0.42\pm0.01$ &  $-0.61\pm0.01$ &   $-1.09\pm0.01$\\
 976\tablefootmark{a}  &  		&   34.745177872  &  57.147424098  & 11.44 &   $0.30\pm0.10$ &  $-0.86\pm0.08$ &   $-1.46\pm0.10$\\
 985  &  		&   34.748616212  &  57.146868131  & 12.06 &   $0.44\pm0.02$ &  $-0.69\pm0.01$ &   $-1.20\pm0.02$\\
 517\tablefootmark{c}  &    BD+56\,489	&   34.562901447  &  57.171116919  & 9.91  &   $0.40\pm0.01$ &  $-0.46\pm0.01$ &   $-1.05\pm0.01$\\
 649\tablefootmark{a}  &  		&   34.611701258  &  57.160105140  & 10.96 &   $0.69\pm0.03$ &  $-1.65\pm0.02$ &   $-1.99\pm0.03$\\
 250\tablefootmark{b}  &  		&   34.404346019  &  57.050049236  & 11.28 &   $0.47\pm0.02$ &  $-0.72\pm0.02$ &   $-1.23\pm0.02$\\
 880\tablefootmark{c,d}  &  		&   34.711257131  &  57.143338200  & 12.97 &   $0.36\pm0.02$ &  $-0.78\pm0.02$ &   $-1.29\pm0.03$\\
 893\tablefootmark{a}  &  		&   34.713035669  &  57.143426004  & 11.89 &  $-0.13\pm0.09$ &  $-0.68\pm0.08$ &   $-1.47\pm0.09$\\
907\tablefootmark{c,d}  &  		&   34.720006422  &  57.141842562  & 12.31 &   $0.44\pm0.02$ &  $-0.86\pm0.01$ &   $-1.36\pm0.02$\\
930\tablefootmark{c,d}  &  		&   34.727762502  &  57.160921960  & 12.31 &   $0.41\pm0.01$ &  $-0.76\pm0.01$ &   $-1.47\pm0.01$\\
1020\tablefootmark{d}&&34.757094378&  57.138732134&13.68&	$0.39\pm0.02$&$-0.67\pm0.01$&$-1.30\pm0.02$\\
264\tablefootmark{b}& &34.414176173&  57.095801754&11.15&	$0.89\pm0.02$ &	$-1.51\pm0.02$	&$-1.02\pm0.02 $\\
1162\tablefootmark{d} &      HD~14143	&   34.808086960  &  57.169225979 &6.51 	  &	  $0.41\pm0.02$&  $-0.59\pm0.01$&  $-1.34\pm0.02$\\
1057 & BD+56\,522	&   34.768546646  & 57.135493752  &6.45&       $0.43\pm0.02$&	$-0.67\pm0.01$&	$-1.10\pm0.02$ \\
16   & HD~13854	  	&34.215479367	&57.055240079 &	6.39&      $0.43\pm0.03$&   $-0.61\pm0.02$&  $-1.12\pm0.02$\\
3\tablefootmark{d}    &HD~13841	 	&34.193289327	& 57.029346575	  &7.32 & $0.41\pm0.02$&   $-0.63\pm0.02$&  $-1.40\pm0.02$\\
\bottomrule
\end{longtable}
\tablefoot{ 
The astrometric membership are considered with all values in parallax, pmRA and pmDEC belong to the interval of 1$\sigma$ from the median cluster values. Characteristics of individual stars:\\
\tablefoottext{a}{RUWE>1.4};
\tablefoottext{b}{bad parallax};
\tablefoottext{c}{bad pmRA};
\tablefoottext{d}{bad pmDEC};
\tablefoottext{e}{bad parallax, pmRA and pmDEC}
}
\setlength{\tabcolsep}{3pt}
\begin{longtable}{llccrcrc}
\caption{Astrometric parameters for stars in NGC~884 from DR3.}
\label{tab:gaia884}\\
\toprule
OO&
Other &
 RA (J2000) &
 DEC (J2000)&
 G&
Plx &
 pmRA&
 pmDEC\\
 Number &
Type &
($^\circ$) &
($^\circ$) &
(mag) &
 (mas)&
 (mas/yr)
 &(mas/yr)\\
 \midrule
 \endfirsthead
 \caption[]{continued.}\\
 \toprule
Oo &
Other &
 RA (J2000)&
 DEC (J2000)&
 G&
Plx &
 pmRA&
 pmDEC\\
Number  &
Name &
($^\circ$)  &
($^\circ$)  &
(mag) &
 (mas)
 &(mas/yr)&
(mas/yr) \\
 \midrule
 \endhead
 \midrule
\multicolumn{8}{l}{Continued on next page}
\endfoot
\bottomrule
\endlastfoot
2361\tablefootmark{b}	&HD~14476      &       35.570651174  &$+57.271928743$  &       8.71    &$0.39\pm0.02$  &$-0.83\pm0.01$&$-1.22\pm0.02$\\
2299\tablefootmark{d}	&BD+56\,575    &       35.535703527  &$+57.124548362$  &       9.09    &$0.38\pm0.02$  &$-0.99\pm0.01$&$-1.50\pm0.02$\\
2402\tablefootmark{d}	&BD+56\,582    &       35.594747887  &$+57.284720783$  &       9.40     &$0.43\pm0.02$  &$-0.49\pm0.01$&$-1.78\pm0.01$\\
2235\tablefootmark{c}	&BD+56\,571     &       35.503921683  &$+57.137304633$  &       9.31    &$0.42\pm0.01$  &$-0.53\pm0.01$&$-1.28\pm0.01$\\
2444\tablefootmark{d}	&BD+56\,584    &       35.624372327  &$+57.208005548$  &       9.44    &$0.43\pm0.02$  &$-0.93\pm0.01$&$-1.34\pm0.02$\\
2488	&BD+56\,586    &       35.651231043  &$+57.081770871$  &       9.91    &$0.42\pm0.02$  &$-0.62\pm0.01$&$-1.18\pm0.01$\\
2165\tablefootmark{b}	&BD+56\,566    &       35.470582164  &$+57.166369065$  &       9.72    &$0.42\pm0.02$  &$-0.79\pm0.01$&$-1.17\pm0.02$\\
2246\tablefootmark{b}&BD+56\,572     &       35.511595156  &$+57.140278509$  &       9.88    &$0.41\pm0.02$  &$-0.72\pm0.01$&$-1.13\pm0.02$\\
2227	&HD~14443      &       35.502396807  &$+57.144959279$  &       7.97    &$0.41\pm0.02$  &$-0.60\pm0.02$&$-1.16\pm0.02$\\
2296	&BD+56\,574    &       35.530760712  &$+57.111732782$  &       8.45    &$0.41\pm0.02$  &$-0.66\pm0.01$&$-1.04\pm0.02$\\
2088	&BD+56\,563    &       35.430827256  &$+57.125772817$  &       9.38    &$0.42\pm0.02$  &$-0.64\pm0.01$&$-1.23\pm0.02$\\
2311\tablefootmark{b}	&BD+56\,576    &       35.540456170  &$+57.117301980$  &       9.32    &$0.40\pm0.02$  &$-0.72\pm0.01$&$-1.17\pm0.02$\\
2541	&HD~14520      &       35.681361639  &$+57.086797479$  &       9.16    &$0.41\pm0.02$  &$-0.56\pm0.01$&$-1.22\pm0.01$\\
2262\tablefootmark{b}	&      		&       35.519019149 &$+57.177462308$  &       10.50    &$0.44\pm0.01$  &$-0.74\pm0.01$&$-1.18\pm0.01$\\
2255\tablefootmark{c}	&      		&       35.516267782 &$+57.153840894$  &       10.65   &$0.42\pm0.01$  &$-0.67\pm0.01$&$-1.04\pm0.01$\\
2185\tablefootmark{a}	&      		&       35.481523980 &$+57.099624885$  &       10.91   &$0.47\pm0.25$  &$0.07\pm0.21 $&$-0.08\pm0.25$\\
2242	&      		&       35.510299861 &$+57.155676533$  &       10.89   &$0.43\pm0.01$  &$-0.70\pm0.01$&$-1.19\pm0.01$\\
\end{longtable}
\tablefoot{
The astrometric membership are considered with all values in pmRA and pmDEC belong to the interval of 1$\sigma$ from the median cluster values. Characteristics of individual stars:\\
\tablefoottext{a}{RUWE>1.4};
\tablefoottext{b}{bad pmRA};
\tablefoottext{c}{bad pmDEC};
\tablefoottext{d}{bad pmRA and pmDEC}
}

\begin{table}[h!]
\caption{Astrometric parameters for Be stars from DR3. 
\label{tab:bestars}}
\centering
\begin{tabular}{llrrcc}
\hline
\hline
\noalign{\smallskip}
Oo &
Other &
 G&
 pmRA&
 pmDEC& Membership\\
Number  &
Name &
(mag) &
 (mas/yr)&
(mas/yr) & \\
 \noalign{\smallskip}
\hline
\hline
\noalign{\smallskip}
847 &   BD+56\,511	&   9.02  &   $-0.14\pm0.01$ &   $-1.01\pm0.02$ & NM\\
 717  &  BD+56\,502	&    9.23  &    $ 0.69\pm0.02$  &    $-0.83\pm0.03$ & NM\\
 1268 &  HD~14162		&  9.29  &   $-0.59\pm0.01$ &   $-1.09\pm0.01$ & 1$\sigma$\\
 1261&   BD+56\,534	&   9.35  &    $-1.16\pm0.01$ &   $-1.69\pm0.01$ & NM\\
 2088	&BD+56\,563    &             9.38    &$-0.64\pm0.01$&$-1.23\pm0.02$ & 1$\sigma$\\
  2402&BD+56\,582    &            9.40      &$-0.49\pm0.01$&$-1.78\pm0.01$& NM\\
2284\tablefootmark{a}	 & BD+56   573 & 9.46&$-0.35\pm0.02$&$-1.46\pm0.02$&NM\\ 
 922 &  BD+56\,515	&    9.47  &   $-0.78\pm0.01$ &   $-1.20\pm0.01$& 3$\sigma$ \\
 2165\tablefootmark{a}	&BD+56\,566    &              9.72     &$-0.79\pm0.01$&$-1.17\pm0.02$& 3$\sigma$ \\
  846  &   BD+56\,509	&   9.87  &    $-0.68\pm0.01$ &   $-1.08\pm0.01$& 1$\sigma$ \\
  517\tablefootmark{c}  &    BD+56\,489	&    9.91  & $-0.46\pm0.01$ &   $-1.05\pm0.01$ &3$\sigma$ \\
1161\tablefootmark{a}&   BD+56\,529	&  10.00   &  $-0.46\pm0.09$ &   $-1.08\pm0.11$& 3$\sigma$ \\
2649\tablefootmark{b} &GGA  163 & 10.18 &$-0.65\pm0.01$&$-1.24\pm0.02$&3$\sigma$ \\ 
2566\tablefootmark{b} &MWC  711  & 10.31&
$-0.66\pm0.01$&$-1.24\pm0.02$&3$\sigma$\\ 
2759\tablefootmark{b} &MWC  448  & 10.47&
$-0.58\pm0.01$&$-1.05\pm0.01$&3$\sigma$\\ 
2262&      		&           10.50     &$-0.74\pm0.01$&$-1.18\pm0.01$&3$\sigma$\\
 \noalign{\smallskip}
\hline
\noalign{\smallskip}
2242	&      		&           10.89    &$-0.70\pm0.01$&$-1.19\pm0.01$& 1$\sigma$\\
 1438 &  		&   10.93 &   $-0.61\pm0.01$ &   $-1.09\pm0.01$& 1$\sigma$\\
 1282 &   GGA\,156	&   10.94 &  $-0.58\pm0.01$ &   $-1.10\pm0.02$& 1$\sigma$\\
2091 & GGA  162  & 11.50&
  $-0.76\pm0.03$ &$-1.16\pm0.04$&1$\sigma$\\ 
1278 & & 11.58  &
$-0.50\pm0.01$&$-1.11\pm0.02$  & 3$\sigma$ \\
1977 & GGA  161   & 12.10&$-0.79\pm0.01$ 
& $-1.24\pm0.01$        &3$\sigma$ \\ 
1114 & GGA  154   & 12.31 &$-0.70\pm0.05$
&$-0.93\pm0.06$       &3$\sigma$\\
1058 & & 13.52 &$-0.61\pm0.01$ 
&$-1.27\pm0.01$       &3$\sigma$\\
1912\tablefootmark{b}  & & 13.79 &$-0.44\pm0.01$ 
&$-1.18\pm0.01$           &3$\sigma$\\
 \noalign{\smallskip}
\hline
 \end{tabular}
\end{table}
\tablefoot{
Be stars within the tidal radius of NGC~869 or NGC~884. Characteristics of individual stars:\\
\tablefoottext{a}{Poor astrometric data.};
\tablefoottext{b}{Just outside the tidal radius of NGC~884.}
}

\section{Additional figures}
\twocolumn

  \begin{figure*}
   \centering
    \includegraphics[width=\columnwidth]{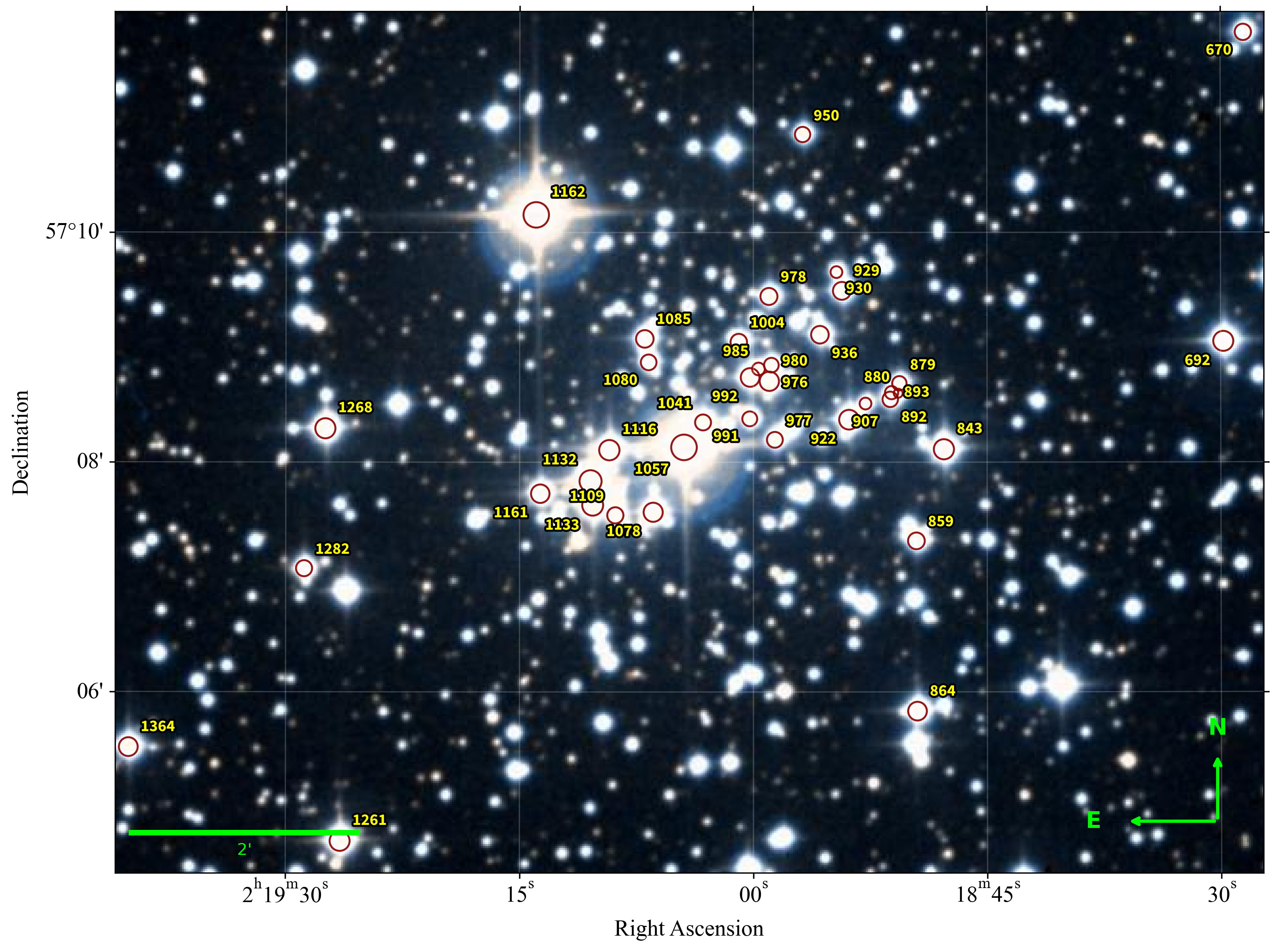}
    \includegraphics[width=\columnwidth]{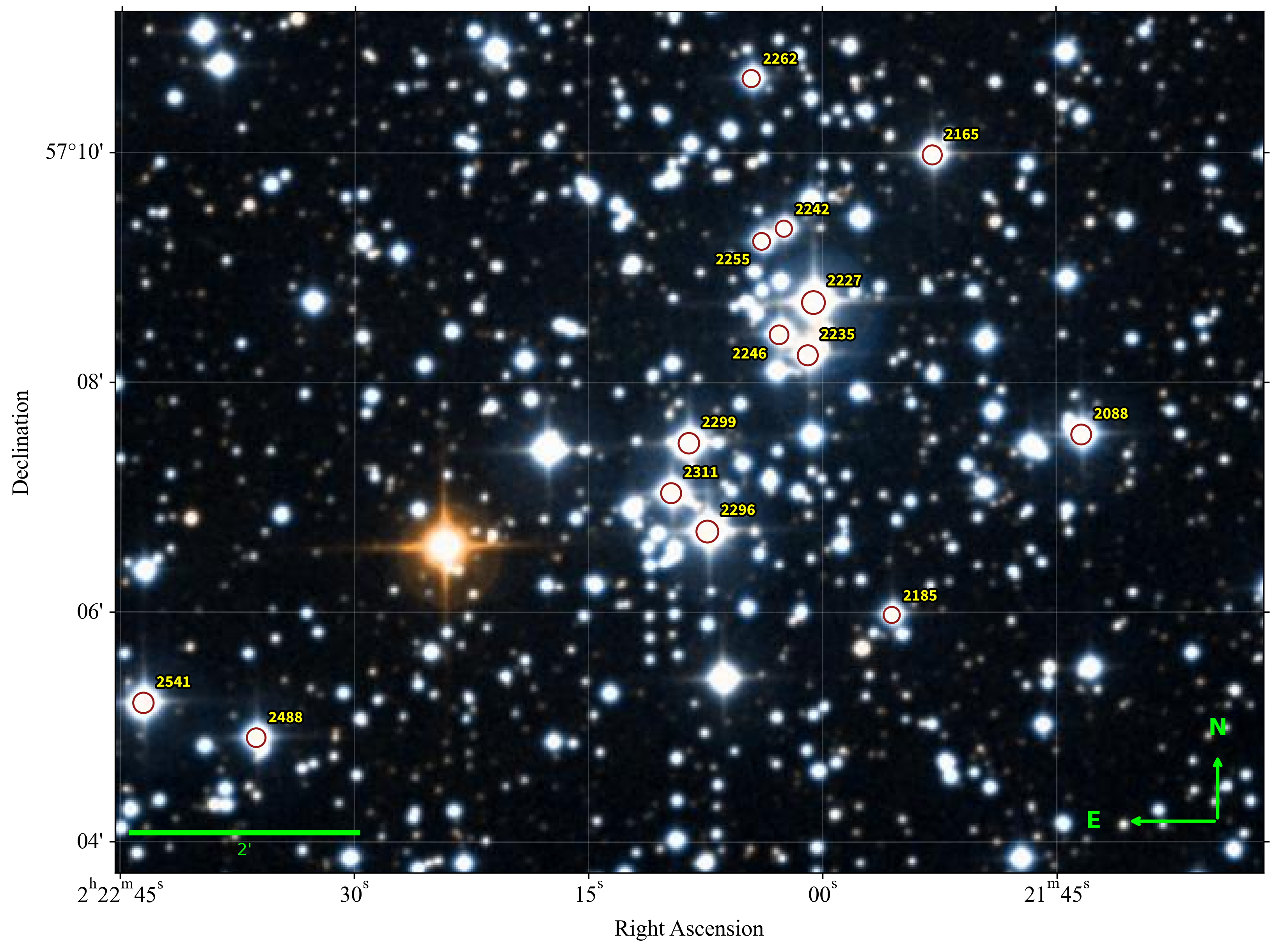}
      \caption{Maps of the core regions for NGC~869 (left) and NGC~884 (right), with all the stars observed marked. The bright red star is the RSG RS~Per, presumably the most evolved member of NGC~884. }
         \label{fig:coremaps}
   \end{figure*}
   
  \begin{figure*}
   \centering
    \includegraphics[width=\columnwidth]{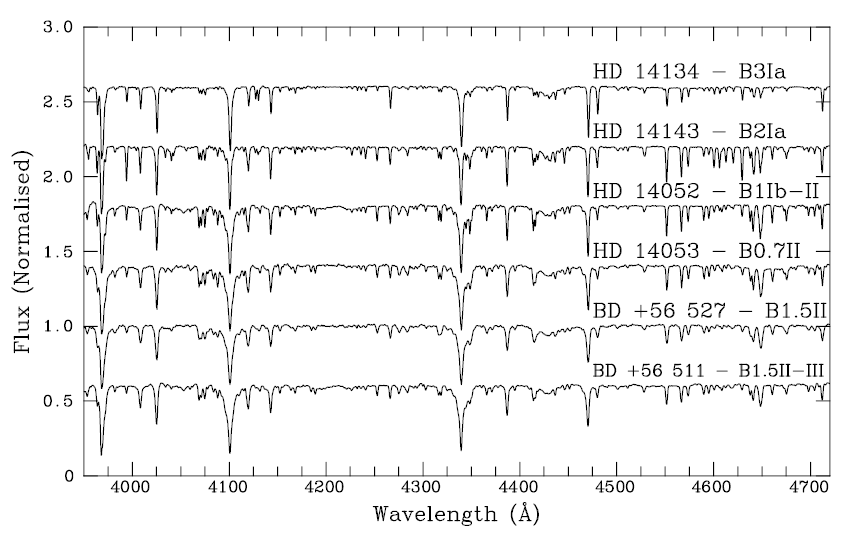}
    \includegraphics[width=\columnwidth]{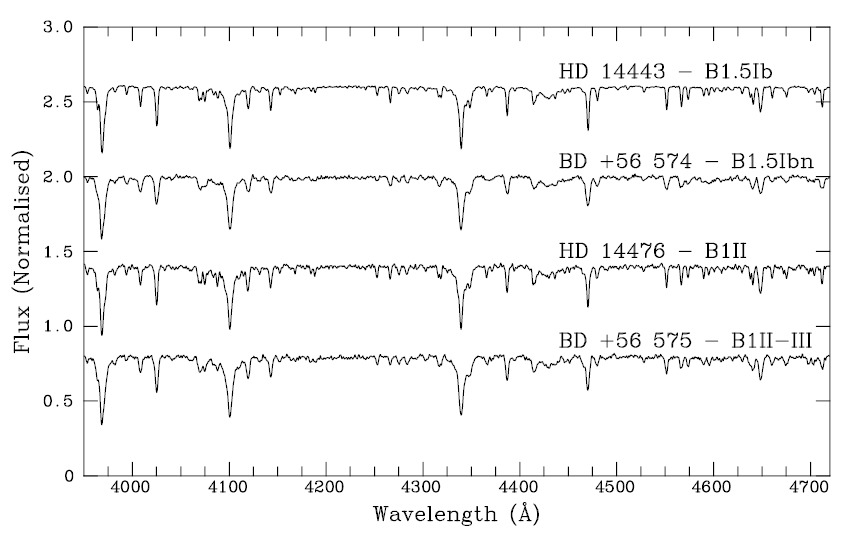}
      \caption{Classification spectra of the most luminous stars in the Double Cluster. \textit{Left: } Stars associated with NGC~869. BD +56 511 is a mild Be star, although this is not obvious in the blue spectrum displayed here. This star is intermediate between B1 and B1.5. \textit{Right: }Stars associated with NGC~884. }
         \label{fig:luminous}
   \end{figure*}

   \begin{figure*}
   \centering
    \includegraphics[width=\columnwidth]{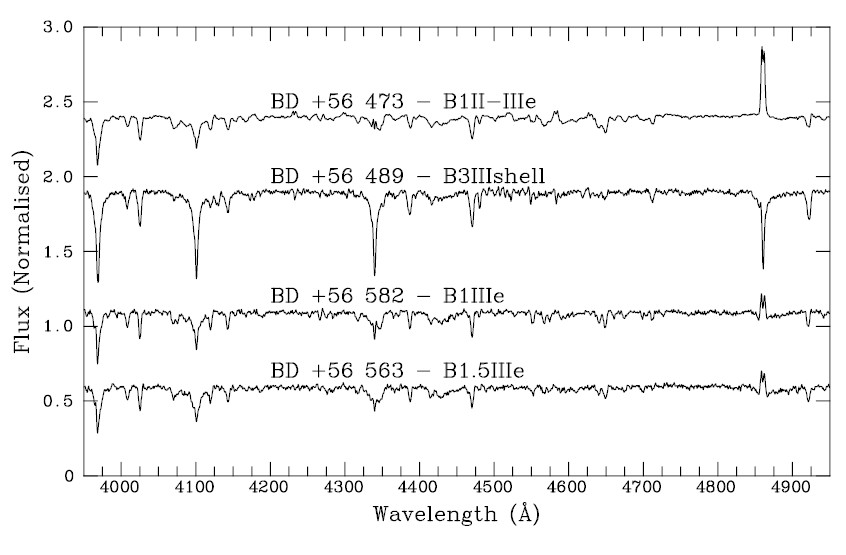}
    \includegraphics[width=\columnwidth]{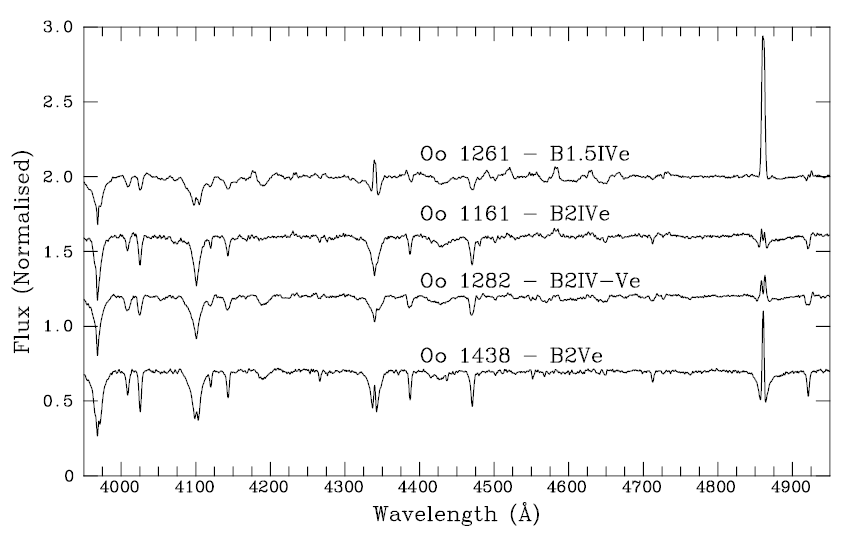}
      \caption{Blue spectra of some Be stars in the Double Cluster. \textit{Left: } Luminous Be stars. \textit{Right: } Less luminous Be stars. Depending on the intensity of the veiling and the inclination angle, emission and absorption features display varied shapes. The absorption feature seen in many spectra around 4190\,\AA\ is an artefact introduced by flat-fielding. }
         \label{fig:bes}
   \end{figure*}

   \begin{figure*}
   \sidecaption
   \includegraphics[width=12cm]{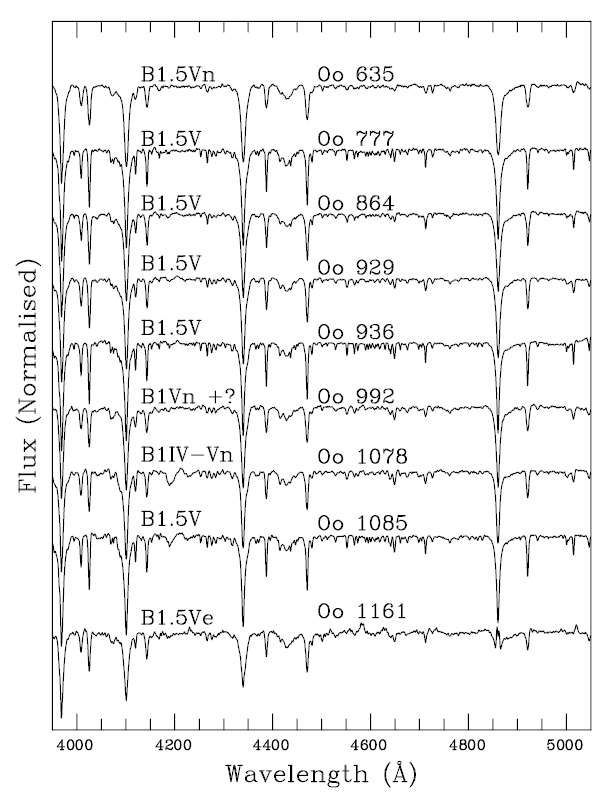}
      \caption{Blue spectra of the earliest dwarfs around the core of NGC~869, defining the cluster turn-off. Most of them correspond to spectral type B1.5\,V. The diversity of morphologies is partly due to rotational velocity, with fast rotators marked with the "n" qualifier, but also to the intrinsic width of spectral subtypes, with Oo~936 and Oo~1085 close to B1\,V, while Oo~864 is very close to B2\,V. Note that some spectra present an artefact introduced by flat-fielding that appears as an  absorption feature around 4190\,\AA.}
         \label{fig:to_h}
   \end{figure*}

   \begin{figure*}
    \sidecaption
\includegraphics[width=12 cm]{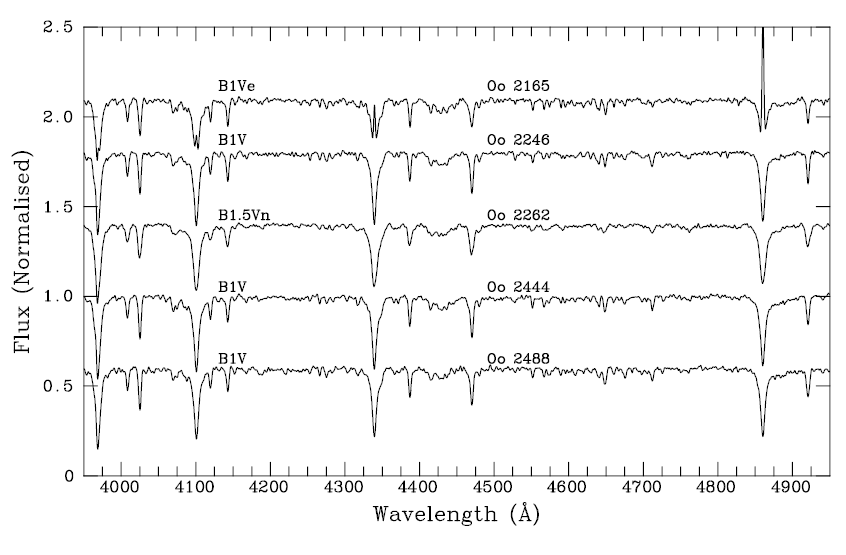}
      \caption{Blue spectra of the earliest dwarfs around the core of NGC~884, defining the cluster turn-off. Most of them correspond to spectral type B1\,V.}
         \label{fig:to_chi}
   \end{figure*}

   \begin{figure*}
   \sidecaption
   \includegraphics[width=6cm]{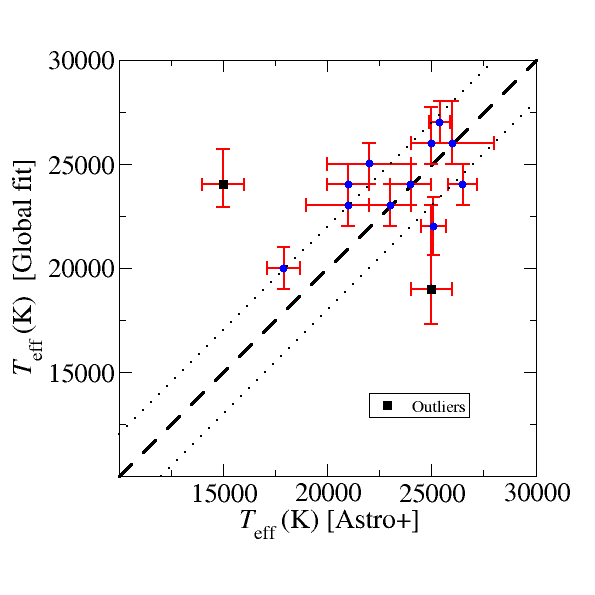}
   \includegraphics[width=6cm]{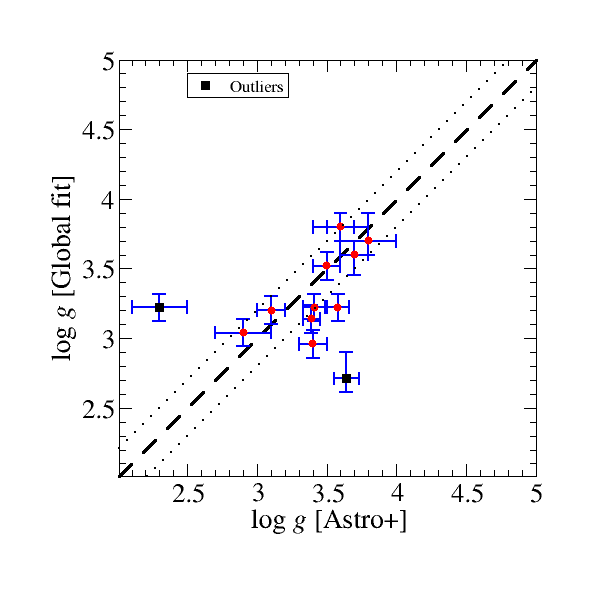}
\caption{Left: Comparison of effective temperature values for stars in NGC~869 obtained with two different codes. The vertical axis represents values obtained using the method based on a global fit of the spectrum to the model \citep{castro2012}. The horizontal axis represents values obtained using the \textit{HiLineThere} algorithm implemented in the Astro+ database \citep{Rubke2026}. The dashed line represents a one-to-one relation. The dotted lines are separated by 2000~K. Two fast rotators (846 and 922) significantly deviate from the bulk of stars. They are marked with a black square in the plot, but note that other stars with similarly high rotational velocities (e.g. 1132) give consistent parameters.
Right: As in the left panel, but now for logarithm of effective gravity. The dotted lines are separated by 0.2~dex. Again, 846 and 922 appear as outliers (see the text for a discussion).}
\label{fig:comparisonTg} 
\end{figure*}

\begin{figure*}
\resizebox{\columnwidth}{!}{\includegraphics[angle=0]{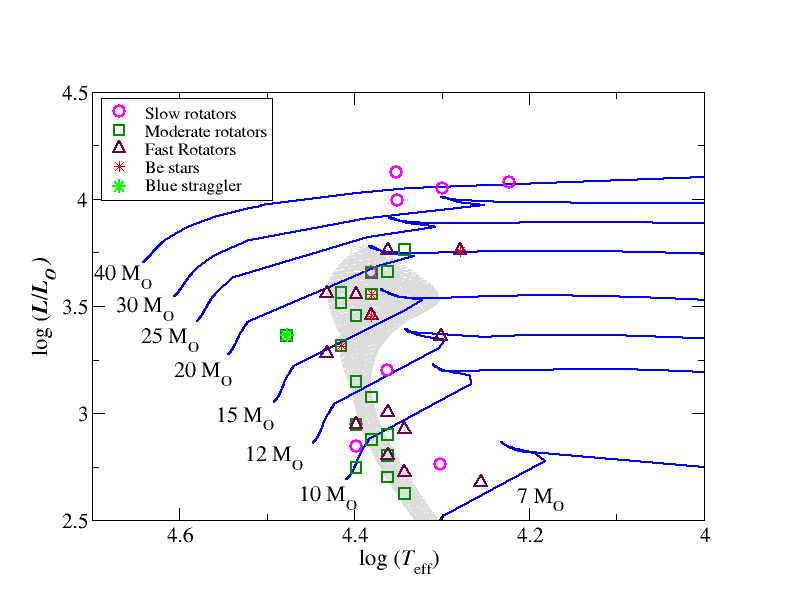}}
\resizebox{\columnwidth}{!}{\includegraphics[angle=0]{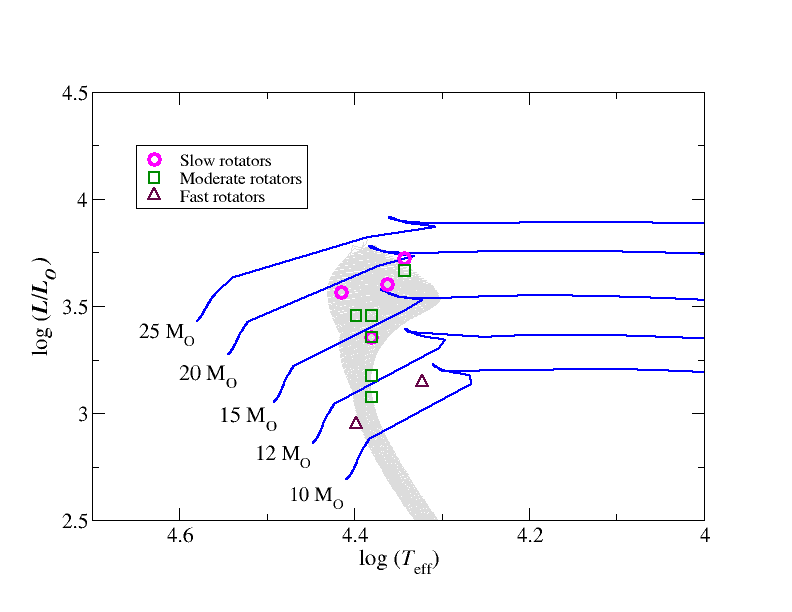}}
\caption{ Spectroscopic HRD for stars in NGC~869 (left) and for stars in NGC~884 (right). Evolutionary tracks and isochrones as in Fig.~\ref{fig:brottdiags}, but now the symbols indicate the measured rotational speed of the stars. Circles mark slow rotators ($v_{\mathrm{rot}} < 100\:\mathrm{km}\,\mathrm{s}^{-1}$); squares indicate stars with moderate rotation (up to $v_{\mathrm{rot}} = 200\:\mathrm{km}\,\mathrm{s}^{-1}$) and triangles are fast rotators ($v_{\mathrm{rot}} > 200\:\mathrm{km}\,\mathrm{s}^{-1}$), which appear filled if they are also Be stars. The only blue straggler is marked as a filled circle, since it belongs to our moderate-rotation category. Note that there may be more than a star at a given location, presenting the same atmospheric parameters, but different rotational velocities.  }   
\label{fig:rotators} 
\end{figure*}


\end{appendix}
\end{document}